\documentclass[a4paper,11pt]{article}
\pdfoutput=1 

\usepackage{jcappub} 

\usepackage[T1]{fontenc} 

\title{\boldmath Null Geodesics and Observables around Kerr-Sen Black Hole}

\author[a]{Rashmi Uniyal}
\author[b]{Hemwati Nandan}
\author[c]{K. D. Purohit}

\affiliation[a]{Department of Physics, \\Government Degree College, \\
Narendranagar-249 175, Uttarakhand India.}
\affiliation[b]{Department of Physics, \\Gurukula Kangri Vishwavidyalaya,\\
Haridwar-249 404, Uttarakhand India.}
\affiliation[c]{Department of Physics, \\Hemwati Nandan Bahuguna Garhwal University,\\
Srinagar Garhwal-246 174, Uttarakhand India.}

\emailAdd{rashmiuniyal001@gmail.com}
\emailAdd{hnandan@associates.iucaa.in}
\emailAdd{kdpurohit@rediffmail.com}

\abstract{We investigate the geodesic motion in the background of Kerr-Sen Black Hole arising in the heterotic string theory.
The nature of effective potential is discussed in radial as well as latitudinal direction. 
A special class of spherical photon orbits is obtained along with the expression for the turning point for radial photons. 
Dependence of photon motion within this class of solution is discussed explicitly in view of the different Black Hole parameters.
We have discussed the allowed regions for geodesic motion of massless test particles around Kerr-Sen Black Hole in more generalised way by including non-equatorial motion of the photons into the account.
The conditions for different types of possible orbits are discussed with specific parameter values along with the corresponding orbit structure. 
No terminating orbits are possible for photons due to non-zero Black Hole charge. 
Observables on the angular plane (viz. bending of light and perihelion precession for massive test particles) are analysed as special cases. 
We have also calculated the rotation and mass parameters for Kerr-Sen Black Hole in terms of the red/blue shifts of the photons in circular and equatorial orbits emitted by the massive test particles which represent stars or other probable sources of photons.}

\begin{document}
\maketitle
\flushbottom
\section{Introduction}

In the low energy limit of string field theory, the Kerr-Sen Black Hole (KSBH) arises as a dilaton-axion generalization of the classical Kerr solution in general theory of relativity (GR) \citep{Sen:1992ua}. 
It has the physical properties similar to the Black Holes (BHs) arising in Einstein-Maxwell theory, but still those can be distinguished in several
aspects \citep{Sen:1992ua, Blaga:2001wt}. 
Most of the Black Hole (BH) solutions in string theory are characterized by one or more charges associated with Yang-Mills fields or the anti-symmetric tensor gauge field. 
The KSBH solution and Gibbons-Maeda (GM) BH/brane solution in a low-energy limit of the heterotic string theory have been known as analytical solutions, which couple to the dilaton and gauge fields \citep{Pradhan:2015yea}.\\ 
\noindent
Recently, the capturing and scattering of photons
in the background of KSBH and Kerr-Newmann BH is studied \citep{Hioki:2008zw} and the evaporation process of these BHs is also investigated in \citep{Koga:1995bs}. 
In addition to these, the hidden conformal symmetries of KSBH is studied in \citep{Ghezelbash:2012qn} and the instability of bound state charged massive scalar fields in such BH spacetime background is discussed in detail in \citep{Furuhashi:2004jk,Siahaan:2015xna}. 
The Conformal Field Theory (CFT) holographic dual for the scattering process in the background of these BHs \citep{Ghezelbash:2014aqa} is also investigated.
In particular, in \citep{Blaga:2001wt}, the radial geodesics around a KSBH are studied with some interesting results. Various important issues such as properties of shadow recasted by KSBH \citep{Dastan:2016bfy}, cosmic censorship conjecture in KSBH \citep{Gwak:2016gwj}, Hawking-temperature and entropy of KSBH \citep{Khani:2014ica, Khani:2013vfa}are discussed in recent past.
Further, the increasing amount of evidence that various galaxies
contain a supermassive BH at their center \citep{Begelman:2003, Shen:2005, Ghez:2008, Morris:2012}, motivated researchers to develop a theoretical approach to obtain the mass
and rotation parameter of a rotating BH in terms of
the redshift $z_{red}$ and blueshift $z_{blue}$ of photons emitted by massive test particles travelling around them along geodesics and the radius of their orbits \citep{Herrera-Aguilar:2015kea, Becerril:2016qxf}.\newpage	
\noindent 
Motivated from the studies on motion of massless and
massive test particles in the background of various BHs in GR \citep{Kagramanova:2010, Grunau:2011, Kagramanova:2012, Grunau:2012, Grunau:2013, Uniyal:2015a, Dasgupta:2012, Teo:2003, Hackmann:2010tqa} and other alternative theories of
gravity \citep{Soroushfar:2015wqa, Uniyal:2014, Dasgupta:2009, Fernando:2012, Olivares:2013, Kuniyal:2015uta, Uniyal:2015sta, Uniyal:thesis, Soroushfar:2016yea}, we make an attempt to understand the null geodesics around a KSBH in detail. 
The present paper is organised as follows. 
A brief introduction to the KSBH spacetime is presented in the next section. 
The geodesic equations and effective potential for incoming test particles is discussed in third section. 
Fourth section presents the discussion on the motion of massless test particles (i.e. photons) around KSBH in which the conditions for the presence of spherical orbits for photons alongwith the corresponding orbit structure is analysed.
In fifth section, few observables for test particles, in the equatorial plane such as bending of light, perihelion precession for massive test particles and red/blue shift of photons emitted by massive test particles are also calculated for a distant observer. Further, these calculated red/blue shifts are used to calculate spin and mass parameters of KSBH theoretically. 
Finally, the key results are summarised in the last section alongwith some future directions.
\section{Kerr Sen BH spacetime}
\noindent
The KSBH spacetime is described by the following $4$D effective action:
\begin{eqnarray}
  S=-\int d^{4}x\sqrt{-\mathcal{G}}e^{-\Phi}\bigg(-\mathcal{R}
       +\frac{1}{12}\mathcal{H}^{2}
       -\mathcal{G}^{\mu\nu}\partial_{\mu}\Phi\partial_{\nu}\Phi
       +\frac{1}{8}\mathcal{F}^{2}\bigg),
       \label{action}
\end{eqnarray}
where $\Phi$ is the dilaton field and $\mathcal{R}$ is the scalar
curvature,
$\mathcal{F}^{2}=\mathcal{F}_{\mu\nu}\mathcal{F}^{\mu\nu}$ with the field strength
$\mathcal{F}_{\mu\nu}=\partial_{\mu}\mathcal{A}_{\nu}-\partial_{\nu}\mathcal{A}_{\mu}$
corresponds to the Maxwell field
$\mathcal{A}_{\mu}$, and
$\mathcal{H}^{2}=\mathcal{H}_{\mu\nu\rho}\mathcal{H}^{\mu\nu\rho}$
with $\mathcal{H}_{\mu\nu\rho}$ given by
\begin{eqnarray}
  \mathcal{H}_{\mu\nu\rho}&=&\partial_{\mu}\mathcal{B}_{\nu\rho}
                 +\partial_{\nu}\mathcal{B}_{\rho\mu}
                 +\partial_{\rho}\mathcal{B}_{\mu\nu}
                 -\frac{1}{4}\bigg(\mathcal{A}_{\mu}\mathcal{F}_{\nu\rho}
                 +\mathcal{A}_{\nu}\mathcal{F}_{\rho\mu}
                 +\mathcal{A}_{\rho}\mathcal{F}_{\mu\nu}\bigg),\label{Hmunurho}
\end{eqnarray}
where the last term in Eq.(\ref{Hmunurho}) is the gauge
Chern-Simons term however $\mathcal{G}_{\mu\nu}$ as appeared in
Eq.(\ref{action}) are the covariant components of the metric in the
string frame, which are related to the Einstein metric by
$g_{\mu\nu}=e^{-\Phi}\mathcal{G}_{\mu\nu}$. 
The Einstein metric, the
non-vanishing components of $A_\mu$, $B_{\mu\nu}$ and the
dilaton field respectively read as below \citep{Sen:1992ua},
\begin{eqnarray}
  ds^{2}&=&-\bigg(\frac{\Delta-a^{2}\sin^{2}\theta}{\Sigma}\bigg)dt^{2}
         +\frac{\Sigma}{\Delta}dr^{2}
         -\frac{4\mu ar\cosh^{2}\alpha\sin^{2}\theta}{\Sigma}dtd\phi
         +\Sigma d\theta^{2}
         +\frac{\Xi\sin^{2}\theta}{\Sigma}d\phi^{2},\label{metric}~~~~~~\\
  \mathcal{A}_{t}&=&\frac{\mu r\sinh 2\alpha}{\sqrt{2}\Sigma},
  \;\;\;\mathcal{A}_{\phi}=\frac{\mu\,a\,r\sinh 2\alpha\sin^{2}\theta}{\sqrt{2}\Sigma},\\
  \mathcal{B}_{t\phi}&=&\frac{2a^{2}\mu r\sin^{2}\theta\sinh^{2}\alpha}{\Sigma},
  \;\Phi=-\frac{1}{2}\ln \frac{\Sigma}{r^{2}+a^{2}\cos^{2}\theta},
\end{eqnarray}
where the metric functions are described as,
\begin{eqnarray}
  \Delta&=&r^{2}-2\mu r+a^{2},\\
  \Sigma&=&r^{2}+a^{2}\cos^{2}\theta+2\mu r\sinh^{2}\alpha,\\
  \Xi&=&\bigg(r^{2}+2\mu r\sinh^{2}\alpha+a^{2}\bigg)^{2}
            - a^{2}\Delta\sin^{2}\theta.
\end{eqnarray}
The parameters $\mu$, $\alpha$ and $a$ are related to the physical
mass $M$, charge $Q$ and angular momentum $J$ as follows,
\begin{eqnarray}
  M=\frac{\mu}{2}(1+\cosh 2\alpha),\hspace{5mm}\;\;
  Q=\frac{\mu}{\sqrt{2}}\sinh^{2}2\alpha,\hspace{5mm}\;\;
  J=\frac{a\mu}{2}(1+\cosh 2\alpha)\hspace{0.5mm}.\label{equation}
\end{eqnarray}
Solving Eq.(\ref{equation}), one can obtain,
\begin{eqnarray}
 \sinh^{2}\alpha=\frac{Q^{2}}{2M^{2}-Q^{2}},\hspace{5mm}\;\;
 \mu=M-\frac{Q^{2}}{2M}\hspace{0.5mm}.\label{relation}
\end{eqnarray}
Then the parameters $\alpha$ and $\mu$ in the metric (\ref{metric})
can be eliminated accordingly. 
For a nonextremal BH, there exist two horizons, determined by $\Delta(r)=0$ as,
\begin{eqnarray}
 r_{\pm}=M-\frac{Q^{2}}{2M}\pm\sqrt{\bigg(M-\frac{Q^{2}}{2M}\bigg)^{2}-a^{2}}\hspace{1mm},
 \label{horizons}
\end{eqnarray}
where $r_{+}$ and $r_{-}$ represent the outer and the inner horizons of the BH respectively.
The case of extremal KSBH requires,
\begin{eqnarray}
 Q^{2}=2M(M-a)\hspace{1mm}.
\end{eqnarray}
The respective ranges of the parameters $a$ and $Q$ are bounded as below, 
\begin{eqnarray}
  0\leq a\leq M,\hspace{5mm}\;\;
  0\leq Q\leq \sqrt{2}M\hspace{1mm}\;\;.
\end{eqnarray}
Here, both the parameters $a$ and $Q$ are considered to be positive and for an extremal KSBH, the two horizons coincide with each other and the degenerate horizon locates at $r_{\text{ex}}=a$.
\section{The geodesic equations and effective potential} 
The first integrals from the geodesic equations in case of a 	KSBH spacetime are calculated as given below \citep{Blaga:2001wt, Dastan:2016bfy, Uniyal:thesis, Soroushfar:2016yea},

\begin{eqnarray}
 \Sigma\frac{dt}{d\tau}&=&\frac{\left(r^2+xr+a^2\right)}{\Delta}\left(E\left(r^2+xr+a^2\right)-aL_z\right)-a\left(aE\sin^2\theta-L_z\right)\,,\label{tequation}\\
 \Sigma\frac{dr}{d\tau}&=&\sigma_{r}\sqrt{\Re}\,,\label{Rad}\\
 \Sigma\frac{d\theta}{d\tau}&=&\sigma_{\theta}\sqrt{\Theta}\,,
          \label{thetaequation}\\
 \Sigma\frac{d\phi}{d\tau}
     &=&\frac{a}{\Delta}\bigg(E\left(r^2+xr+a^2\right)-aL_z\bigg)
              -\frac{1}{\sin^2\theta}\left(aE\sin^2\theta-L_z\right)\,,   
     \label{phiequation}
\end{eqnarray}
with,
\begin{eqnarray}
 {\Re}&=&\bigg[E\big(r(r+x)+a^{2}\big)-aL_z\bigg]^{2}
            -\Delta(\delta r(r+x)+\mathcal{K})\,,\nonumber\\
 \Theta&=&\mathcal{K}-\delta a^2\cos^2\theta-
            \frac{1}{\sin^2\theta}\bigg[aE\sin^2\theta-L_z\bigg]^2\,,
            \label{r-theta-function}
\end{eqnarray}
where the proper time $\tau$ is related to some affine parameter $\lambda$ as $\tau=\delta \lambda$, here $\delta=-1$, 0, 1 for spacelike,
null and timelike geodesics respectively. The constants $E$ and $L_z$ are the conserved energy and orbital angular momentum per unit mass which correspond to the Killing fields $\partial_{t}$ and $\partial_{\phi}$ respectively and $x=Q^2/M$. The variable $\mathcal{K}$ is a separation constant.
The sign functions $\sigma_{r}=\pm$ and $\sigma_{\theta}=\pm$ are
independent from each other. 
The motion of test particles is not confined in a plane in case of axially symmetric spacetimes, hence two effective potentials can be introduced separately corresponding to the radial and latitudinal equations of motion. 
Using Eq.(\ref{Rad}), the effective potential corresponding to the radial motion i.e. $V_{eff,r}=E$ can be obtained with the condition $\dot{r}^2=0$ as,
\begin{equation}
V_{eff,r}=\frac{aL_z\pm\sqrt{\Delta\left(\delta r(r+x)+\mathcal{K}\right)}}{r(r+x)+a^2},
\label{r_Eff_Potential}
\end{equation}
where $\dot{r}^2\geq0$ for $E\leq{V^-_{eff,r}}$ and $E\geq{V^+_{eff,r}}$. 
Similarly the effective potential for latitudinal motion of the test particles i.e. $V_{eff,\theta}=E$ can be obtained by using Eq.(\ref{thetaequation}) as,
\begin{equation}
V_{eff,\theta}=\frac{L_z\pm\sqrt{\sin^2\theta\left(\mathcal{K}-\delta a^2\cos^2\theta\right)}}{a\sin^2\theta},
\label{lat_Eff_Potential}
\end{equation}
where $\dot{\theta}^2\geq0$ for ${V^-_{eff,\theta}}\leq$ E $\leq{V^+_{eff,\theta}}$. 
It is worth noticing that the charge arising in the heterotic sting theory has no significant effect on the motion of test particles in the latitudinal direction. 
Hence, the latitudinal motion of test particles around KSBH is similar to that of Kerr Black Hole (KBH) for respective choice of $\delta$, i.e. $-1$ and $0$ for massive and massless test particles respectively.
\section{Motion of Photons around KSBH}
\subsection{Conditions for Spherical Orbits}
\noindent
With the substitution $u=\cos\theta$, $Y={L_z}/E$ and $Z=\left(\mathcal{K}-({L_z}-aE)^2\right)/E^2$, the equation of motion in $\theta$-direction can be transcribed as,
\begin{equation}
\left(\frac{\Sigma}{E}\right)^2\dot{u}^2\,=\,\Theta(u)=\,Z-\left(Z+Y^2-a^2\right)u^2-a^2u^4.
\label{eq:theta-null}
\end{equation}
The physically allowed regions for the geodesic motion of test particles correspond to $\Theta(u)>0$, which are bounded by $\Theta(u)=0$ values.
Different possible solutions corresponding to the different settings of $Z$ values are discussed in greater detail in \citep{Teo:2003} for the motion of photons in latitudinal direction. 
Since the presence of charge in case of KSBH, does not effect the motion of photons in latitudinal direction, the different solutions obtained for photon's latitudinal motion around KBH \citep{Teo:2003} are used here to discuss the geodesic motion in general. 
The radial equation of motion for photons may then be written as,
\begin{equation}
\left(\frac{\Sigma}{E}\right)^2{\dot{r}}^2=\mathcal{R}(r)
\nonumber
\end{equation}
\begin{equation}
=2M\mathcal{S}x-a^2 Z+\left(2M\mathcal{S}+\mathcal{T}x\right)r+\left(x^2+\mathcal{T}\right)r^2+2xr^3+r^4,
\end{equation}
where,
\begin{equation}
\mathcal{S}=\left(a-Y\right)^2+Z,\hspace{3cm}
\mathcal{T}=a^2-Y^2-Z.\nonumber
\label{sph_f_01}
\end{equation}
For spherical photon orbits at constant radius $r$, the conditions $\mathcal{R}(r)=\frac{d\mathcal{R}(r)}{dr}=0$ must hold at this constant value of $r$. On solving these two equations simultaneously, one arrives with two one-parameter classe solutions	 parametrized in terms of $r$ as follows,\vspace{2mm}\\
{\bf\underline{Case (i)}: } $Y=\frac{1}{a}\,\left(r(r+x)+a^2\right)$,   $Z=-\frac{r^2}{a^2}(r+x)^2$;\vspace{2mm}\\
{\bf\underline{Case (ii)}: }
$Y=-\frac{1}{a(r-M)}\left[r^3+(x-3M)r^2+(a^2-3Mx)r+a^2M\right]$,
\begin{equation}
Z=-\frac{r^2(r+x)}{a^2(r-M)^2}\left[r^3+(x-6M)r^2+3M(3M-2x)r-4a^2M+9M^2x	\right].
\label{sph_f_02}
\end{equation}
\begin{figure}[h!]
\centering{\includegraphics[scale=0.45]{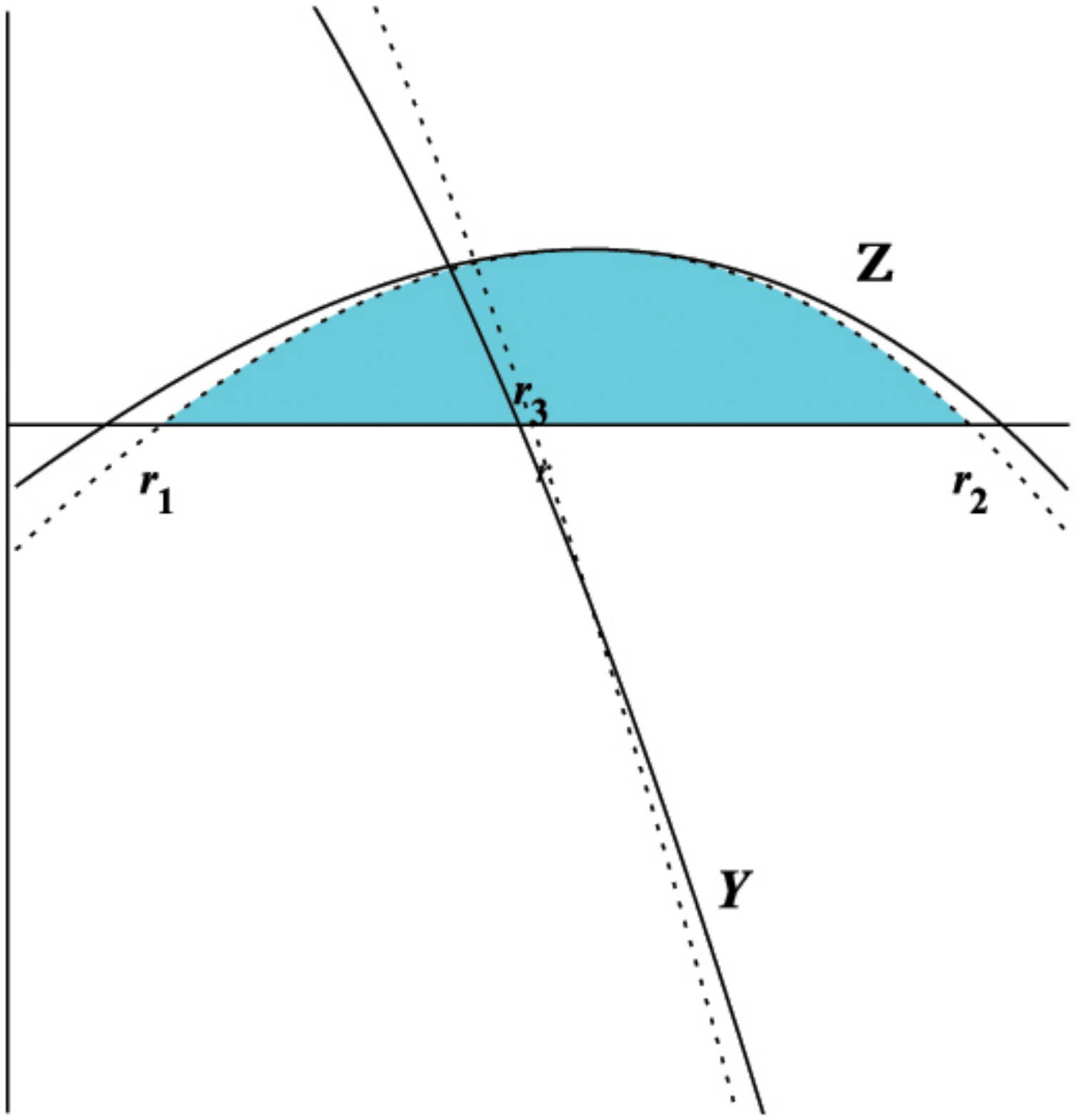}}
\caption{\label{fig:Q-phi-r} Qualitative description for $Z$ and $Y$ with $r$ for solutions of case (ii) in Eq.(\ref{sph_f_02}).}
\end{figure}\\
\noindent 
Among the abovementioned two classes of the solutions, only the class (ii) solutions are physically plausible as class (i) solution violates the necessary condition for the latitudinal motion of photon.
As depicted in Fig.(\ref{fig:Q-phi-r}), the behaviour of $Z$ and $Y$ is qualitatively similar to that of a KBH spacetime. The physical range for spherical photon orbits lies in the range $r_1<r<r_2$, which shrinks in the presence of the charge. 
The value of $Z$ attains its maximum value i.e. $9M(3M+x)$ in between $r_1$ and $r_2$. 
Since $Z$ is related to the angular speed of the photon passing through equatorial plane in $\theta$-direction and for $Z=0$, the photons must lie entirely in the equatorial plane. 
As the value of $Z$ increases, the photon acquires velocity in $\theta$-direction as well. 
At $Z_{max}$, photon's velocity now has no $\phi$-component.
\par\noindent $Y$ is related to the angular momentum of photon about $\phi$-axis. It monotonically decreases in the range $r_1<r<r_2$. 
It vanishes at an intermediate value of $r$ given below, 
\begin{eqnarray}
r_3=M-\frac{x}{3}+2\sqrt{M^2-\frac{a^2}{3}+\frac{Mx}{3}+\frac{x^2}{9}}\cos\left(\frac{1}{3}\arccos\frac{\xi_1}{\xi_2}\right),
\nonumber\\
\xi_1=M(M^2-a^2)+\frac{M^2x}{2}+\frac{a^2x}{6}-\frac{Mx^2}{6}-\frac{x^3}{27},\nonumber\\
\xi_2=\left(M^2-\frac{a^2}{3}+\frac{Mx}{3}+\frac{x^2}{9}\right)^{3/2}.
\end{eqnarray}
$r_3$ corresponds to the photon with \textit{zero angular momentum}. 
At $r=r_3$, the angular momentum of test particle vanishes about $\phi$-axis, but still this point corresponds to some finite value of $Z$-function. 
As explained previously, $Z$-function corresponds to the angular speed of the photon along $\theta$-direction, hence a photon with zero angular momentum also has angular velocity in $\theta$-direction, which is actually an effect of spacetime curvature due to the presence of rotation, also known as {\it frame-dragging-effect}.
\subsection{Latitudinal Motion}
\begin{figure}[h!]
\centerline{
\includegraphics[scale=0.27]{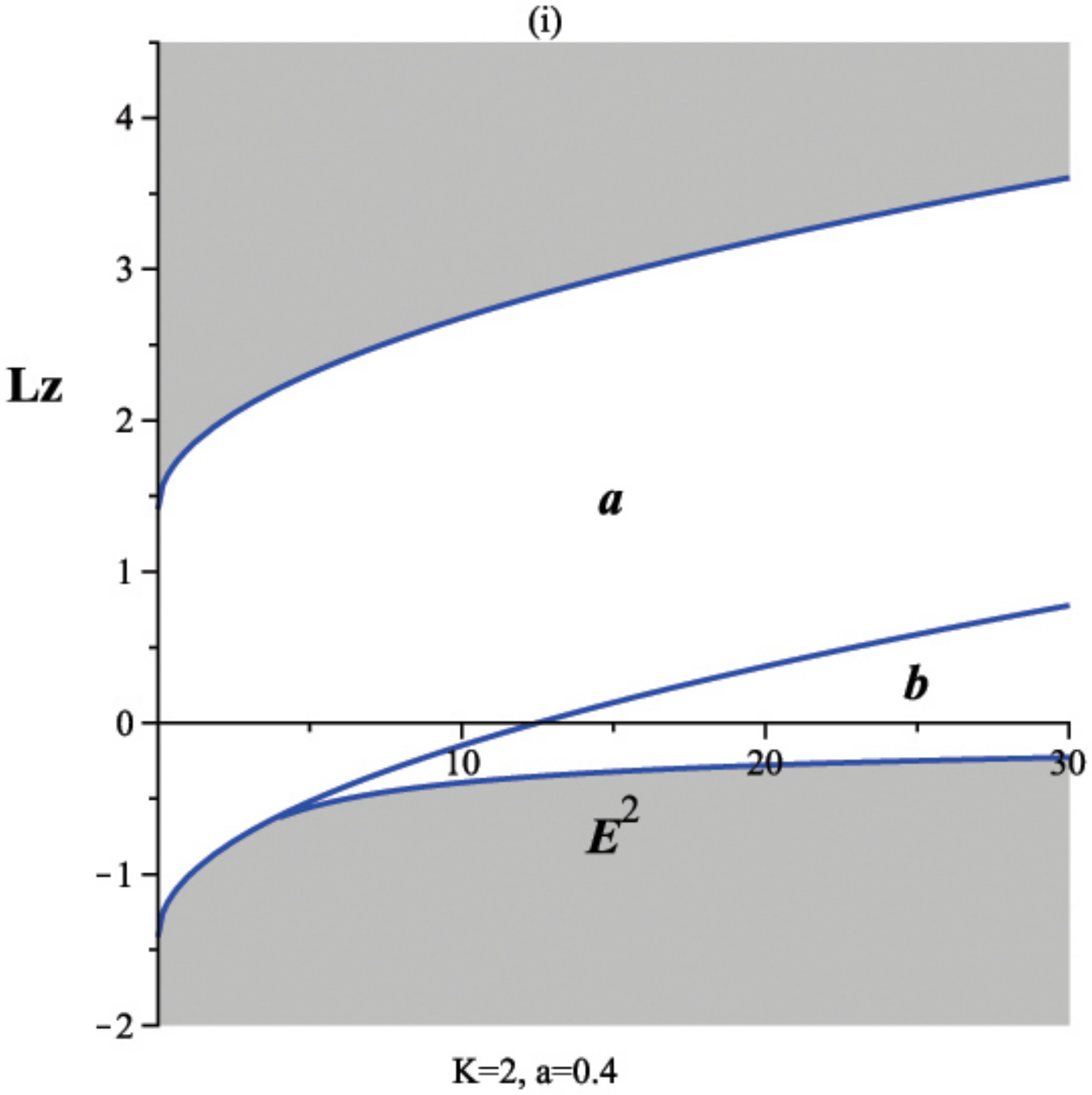}
\includegraphics[scale=0.27]{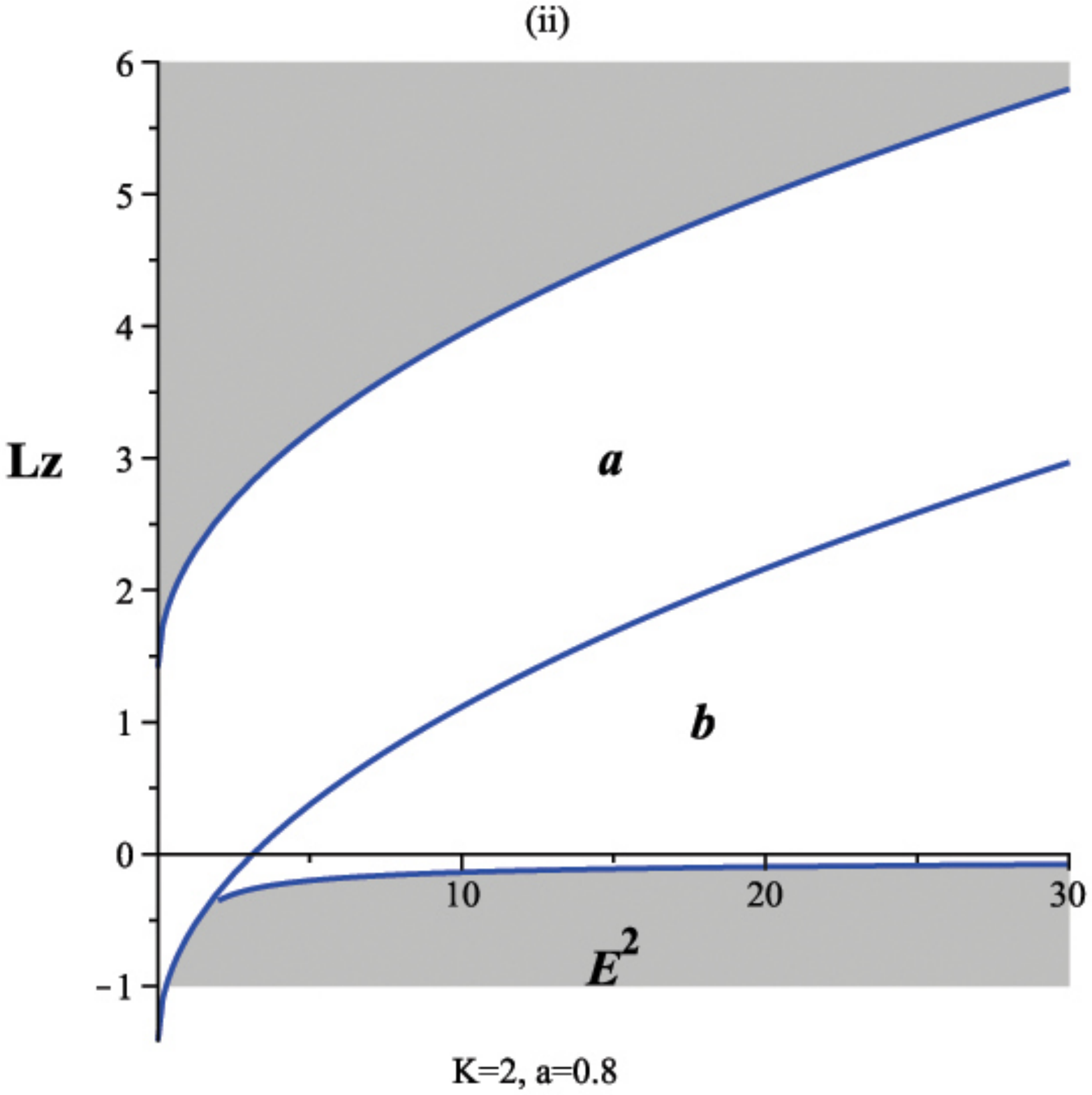}
\includegraphics[scale=0.27]{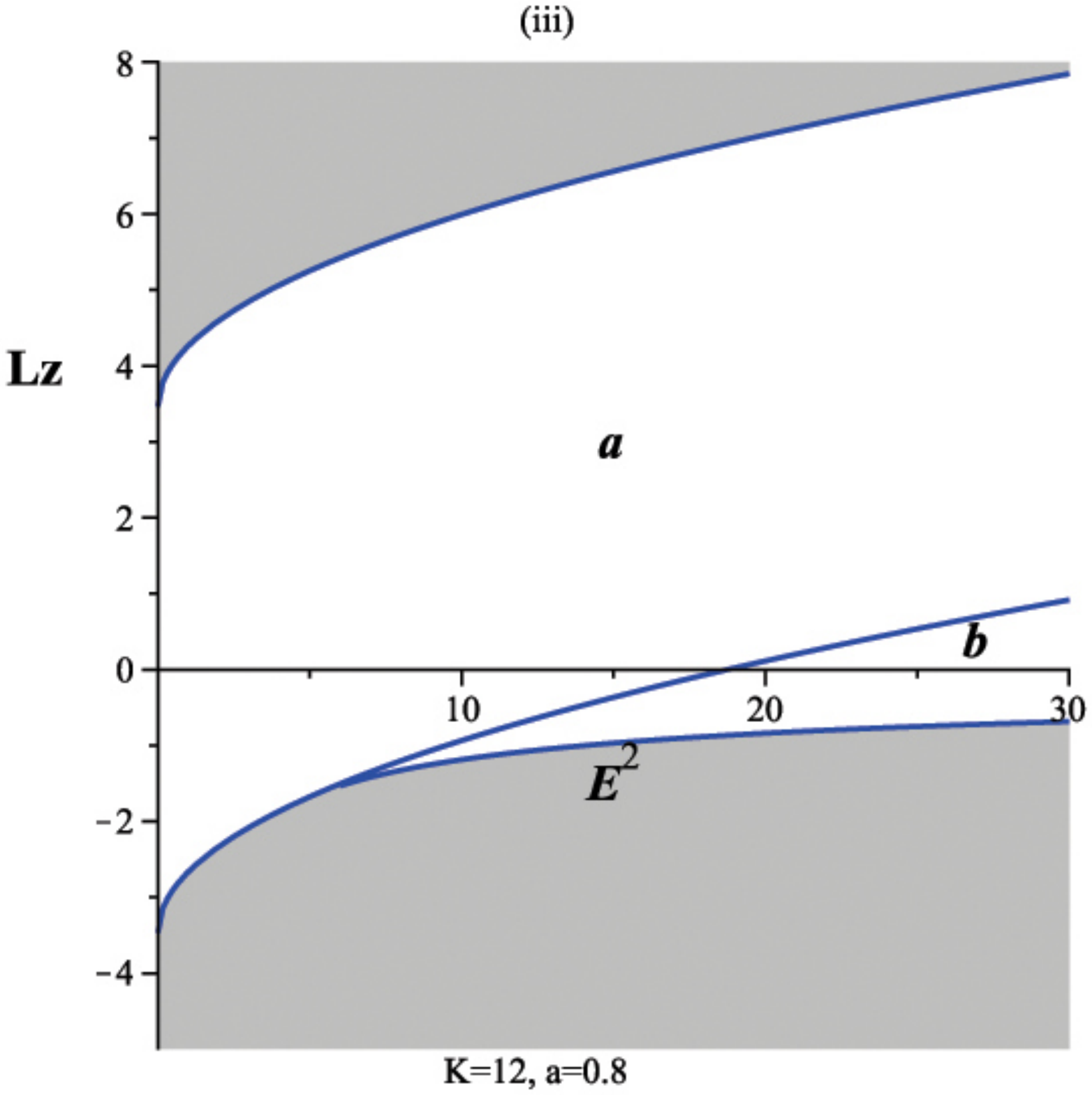}
}
\caption{ Parametric $L_z-E^2$ diagram for $\Theta$. Here shaded regions depict the forbidden areas for geodesic motion of a massless test particle.\label{fig:latitudinal_motion}}
\end{figure}

\noindent 
As geodesic motion is possible for ${\Theta}(\theta)$ $\geq$ $0$, hence the motion of a test particle in latitudinal direction depends on the nature of zeros of the polynomial $\Theta$.
For the determination of the latitudinal motion of photon the function ${\Theta}(\theta)$ is used from Eq.(\ref{phiequation}) with $u=\cos^2\theta$ and $\delta=0$ as,
\begin{equation}
\Theta(u)=\mathcal{K}-\left(a^2E^2(1-u)-2aEL_z+\frac{L_z^2}{(1-u)}\right).
\label{theta-null}
\end{equation}
The number of zeros of $\Theta(\theta)$ changes only if a zero crosses $0$,$1$ or a double zero. 
Further, $u=0$ becomes a zero of $\Theta(u)$, if
\begin{equation}
\Theta(u=0)\,=\,\mathcal{K}-a^2\,E^2+2\,a\,E\,L_z+L_z^2\,=\,0,
\label{theta-null-zero}
\end{equation}
which in turn implies that,
\begin{equation}
L_z=a\,E\pm\sqrt{\mathcal{K}}.
\label{Lat-limit-zero}
\end{equation}
Since $u=1$ is the pole of $\Theta(u)$ for $L_z\neq0$ and it becomes a zero of $\Theta(u)$ only for $L_z=0$.
\begin{equation}
\Theta(u=1,L_z=0)=\mathcal{K}.
\label{theta-null-01}
\end{equation}
On removal of the singularity at $u=1$ (i.e. $\theta$ = 0, $\pi$), the zeros of $\Theta(u)$ are given by the zeros of the function,
\begin{eqnarray}
\Theta(u)=(1-u)\,\mathcal{K}-\left(a\,E-L_z-a\,E\,u\right)^2\nonumber\\
\,\,\,\,\,=-a^2\,E^2\,u+\left(2\,a\,E(a\,E-L_z)-\mathcal{K}\right)\,u+\mathcal{K}-\left(a\,E-L_z\right)^2,
\label{theta-null-02}
\end{eqnarray}
and $\Theta(u)$ has a double zero iff,
\begin{equation}
\Theta(u)=0,\,\,\,\,\frac{d\Theta(u)}{u}=0,
\label{theta-double-zero-condition}
\end{equation}
which reduces to the following simple form,
\begin{equation}
L_z=-\frac{\mathcal{K}}{4\,a\,E}.
\label{theta-null=d}
\end{equation}
\\
\noindent Using conditions given in Eq.(\ref{Lat-limit-zero}), Eq.(\ref{theta-null-02}) and Eq.(\ref{theta-null=d}), the parametric  $L_z-E^2$ diagrams can be plotted as shown in Fig.(\ref{fig:latitudinal_motion}). 
The Fig.(\ref{fig:latitudinal_motion}) depicts the possible regions for geodesic motion. 
In region $a$, $\Theta$ has only one zero and hence geodesics can cross the equatorial plane (i.e. $K>\left(L_z - aE\right)^2$). 
In region $b$, $\Theta$ has two zeros, corresponding to the motion above and below the equatorial plane (i.e. $K<\left(L_z - aE\right)^2$). 
For the motion confined in the equatorial plane $K=\left(L_z - aE\right)^2$). 
For a larger value of BH spin, region $b$ increases in comparison to region $a$ (see Fig.(\ref{fig:latitudinal_motion}(ii))) while for larger $\mathcal{K}$ values, region $a$ becomes larger (see Fig.(\ref{fig:latitudinal_motion}(iii))).Setting $Q$ and $a$ to zero, respectively, one can obtain the maximum
values for them.
\subsection{Radial Motion}
\begin{figure}[h!]
\centerline{
\includegraphics[scale=0.3]{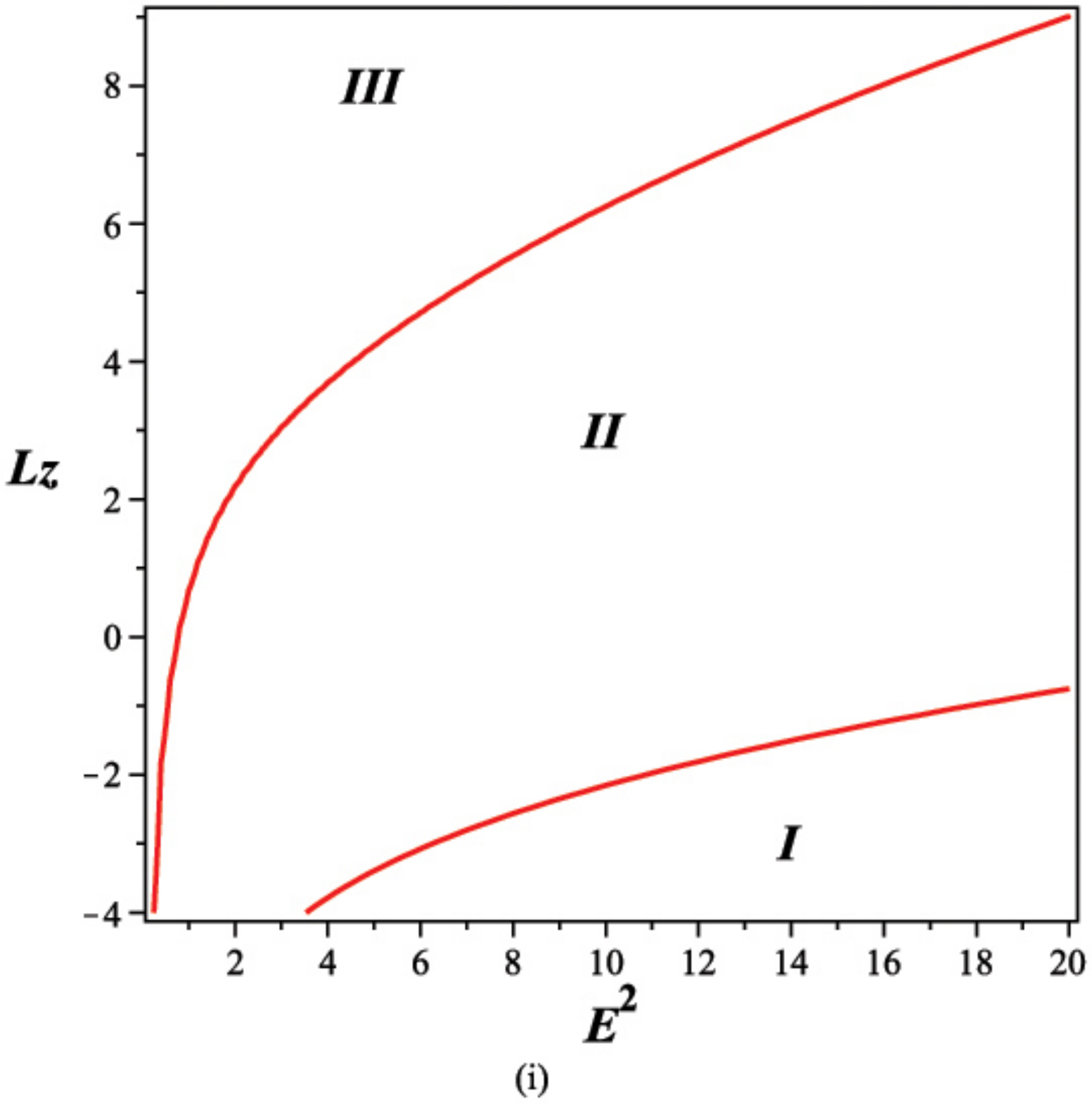}
\includegraphics[scale=0.3]{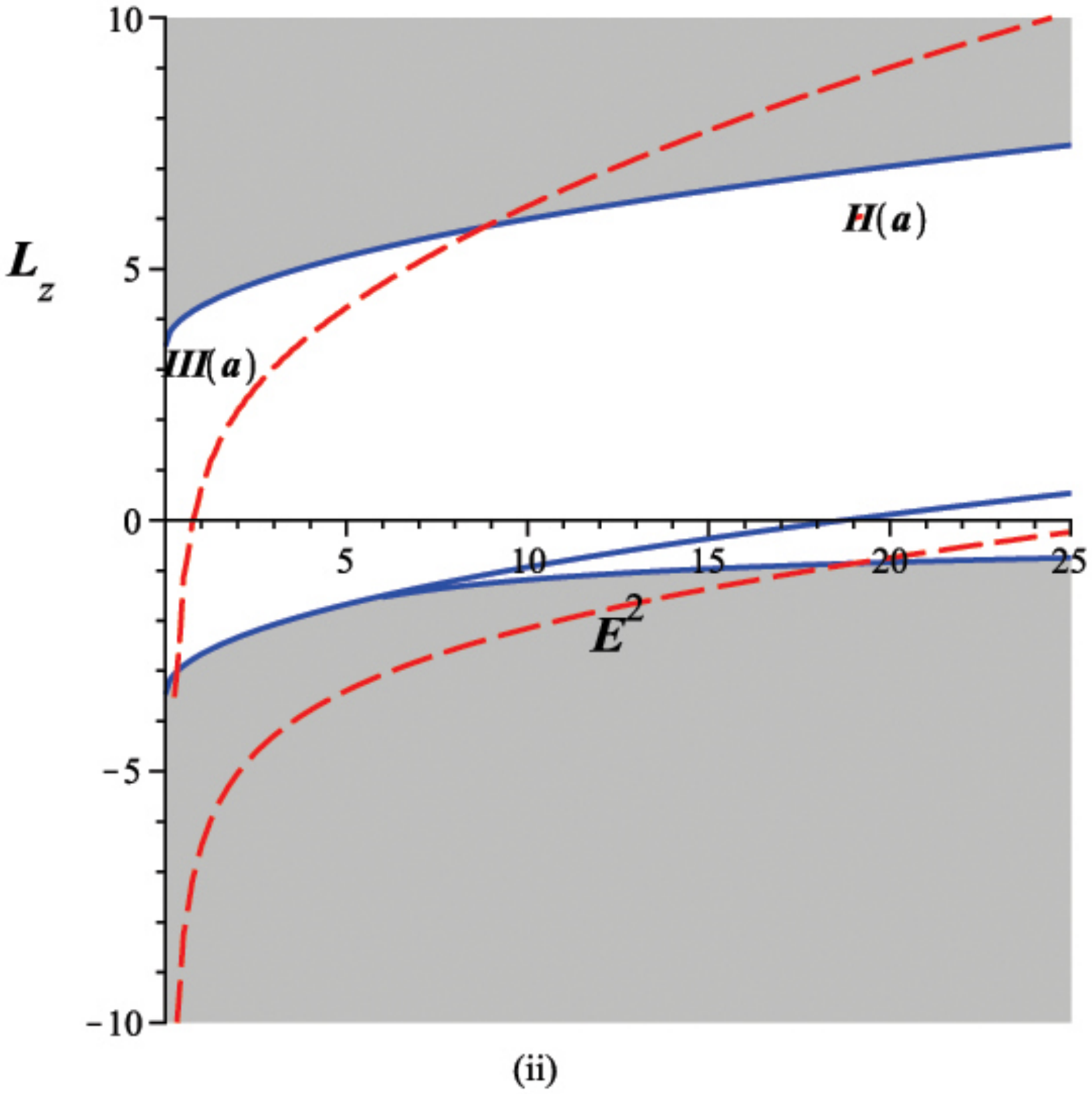}
\includegraphics[scale=0.3]{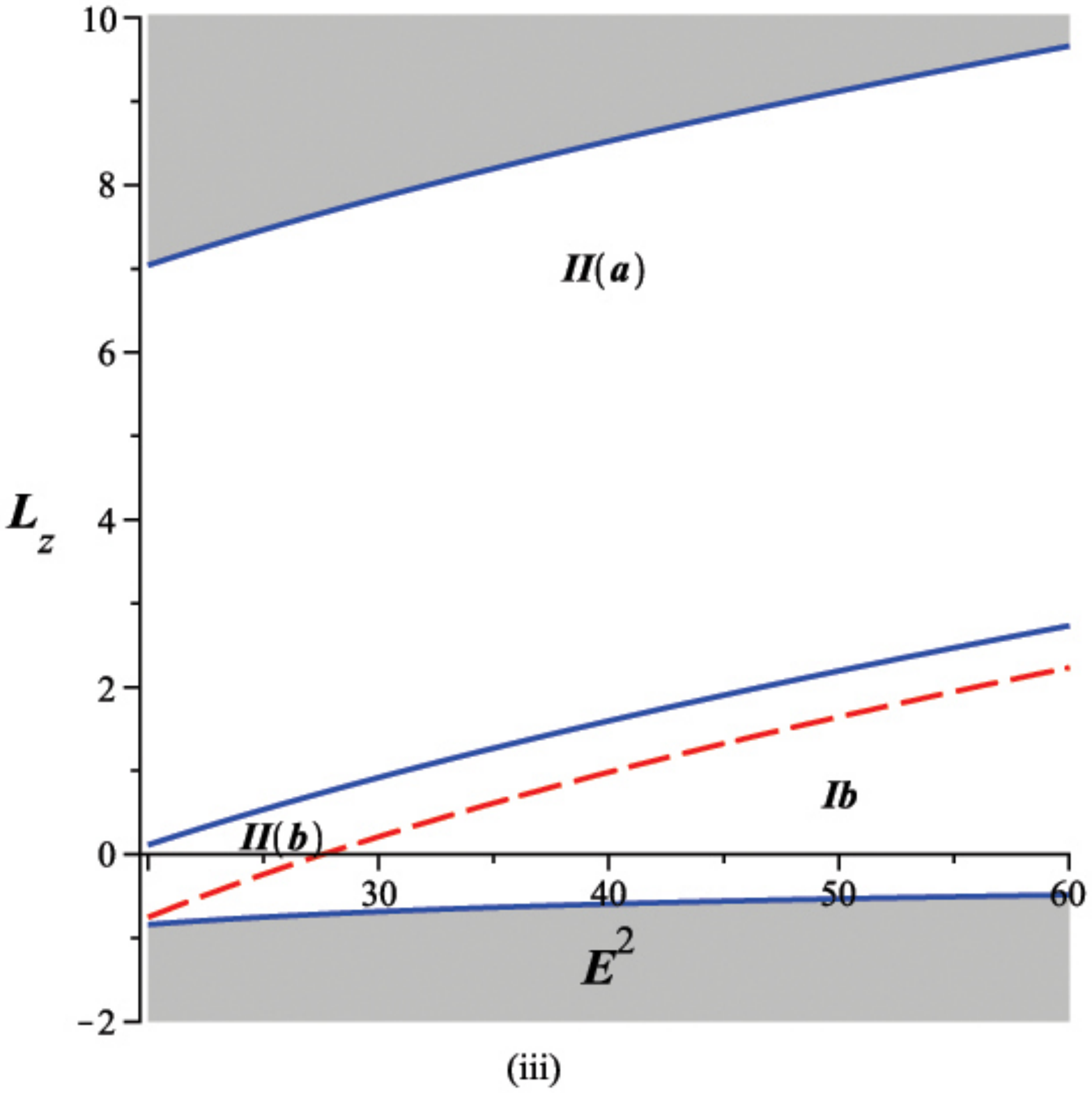}
}
\caption{ Parametric $L_z-E^2$ diagram for radial direction; (i) Regions of $r$-motion; (ii) Combination with $\theta$-motion; (iii) Enlarged view of the region $\bf{b}$; with $a=0.8$, $M=1$, $Q=0.4$, $K=12$. Here solid lines represent the bounderies of $\theta$-motion while dotted-dashed lines represent bounderies of $r$-motion and shaded areas depict the forbidden regions for geodesic motion.\label{fig:radial_motion_null}}
\end{figure}

In order to determine the motion of a test particle in radial direction, one has to analyse the radial function $\Re$ given in Eq.(\ref{r-theta-function}). 
For massless test particles (i.e. with $\delta=0$), the radial function $\Re$ reduces to the following form,
\begin{equation}
\mathcal{R} (r)=\bigg[E\big(r(r+x)+a^{2}\big)-aL_z\bigg]^{2}-\Delta\mathcal{K}.
\label{radial-null-function}
\end{equation}
Now the motion of test particles depends on the nature of zeros of the function $\mathcal{R}$. 
These zeros (also known as turning points) change if a double zero occurs. 
The condition for the occurrence of the double zero is given by,
\begin{equation}
\mathcal{R}(r)=0,\,\,\,\,\,\,\,\frac{d\mathcal{R}(r)}{dr}=0.
\label{radial-doule-zero-condition}
\end{equation}
On solving the Eq.(\ref{radial-doule-zero-condition}) simultaneously, one can have the parametric $L_z-E^2$ representation for massless test particles in radial direction as depicted in Fig.(\ref{fig:radial_motion_null}(i)). 
As geodesic motion is possible if $\mathcal{R}(r)\,\geq$ 0. 
Fig.(\ref{fig:radial_motion_null}(i)) represents the three possible regions corresponding to the different nature of the roots of the function $\mathcal{R}(r)$.
In the most general cases, there may be the following possibilities for the $r$-motion of an arbitrary test particle according to the nature of the roots of $\mathcal{R}(r)$ \citep{Soroushfar:2015wqa, Hackmann:2010tqa}:
\begin{itemize}
\item[(a)] All zeros of $\mathcal{R} (r)$ are complex and $\mathcal{R}(r)\geq 0$ for all $r$ values. Possible orbit type: transit orbit.
\item[(b)] $\mathcal{R} (r)$ has two real zeros $r_1$ and $r_2$ such that $\mathcal{R}(r)\geq 0$ for $r\leq r_1$ and $r_2\leq r$.
\item[(c)] All zeros of $\mathcal{R}(r)$ are real such that, $\mathcal{R}(r)\geq 0$ for ${r_1}\,\leq\,r\,\leq\,{r_2}$ and ${r_3}\,\leq\,r\,\leq\,{r_4}$. 
The possible orbit types may be: two different bound orbits.
\item[(d)] All zeros of $\mathcal{R}(r)$ are real such that, $\mathcal{R}\geq\,0$ for $r\,\leq\,{r_1}$, ${r_2}\,\leq\,r\,\leq\,{r_3}$ and ${r_4}\,\leq\,r$.
\end{itemize}
\begin{figure}[h!]
\centerline{
\includegraphics[width=7cm,height=7cm]{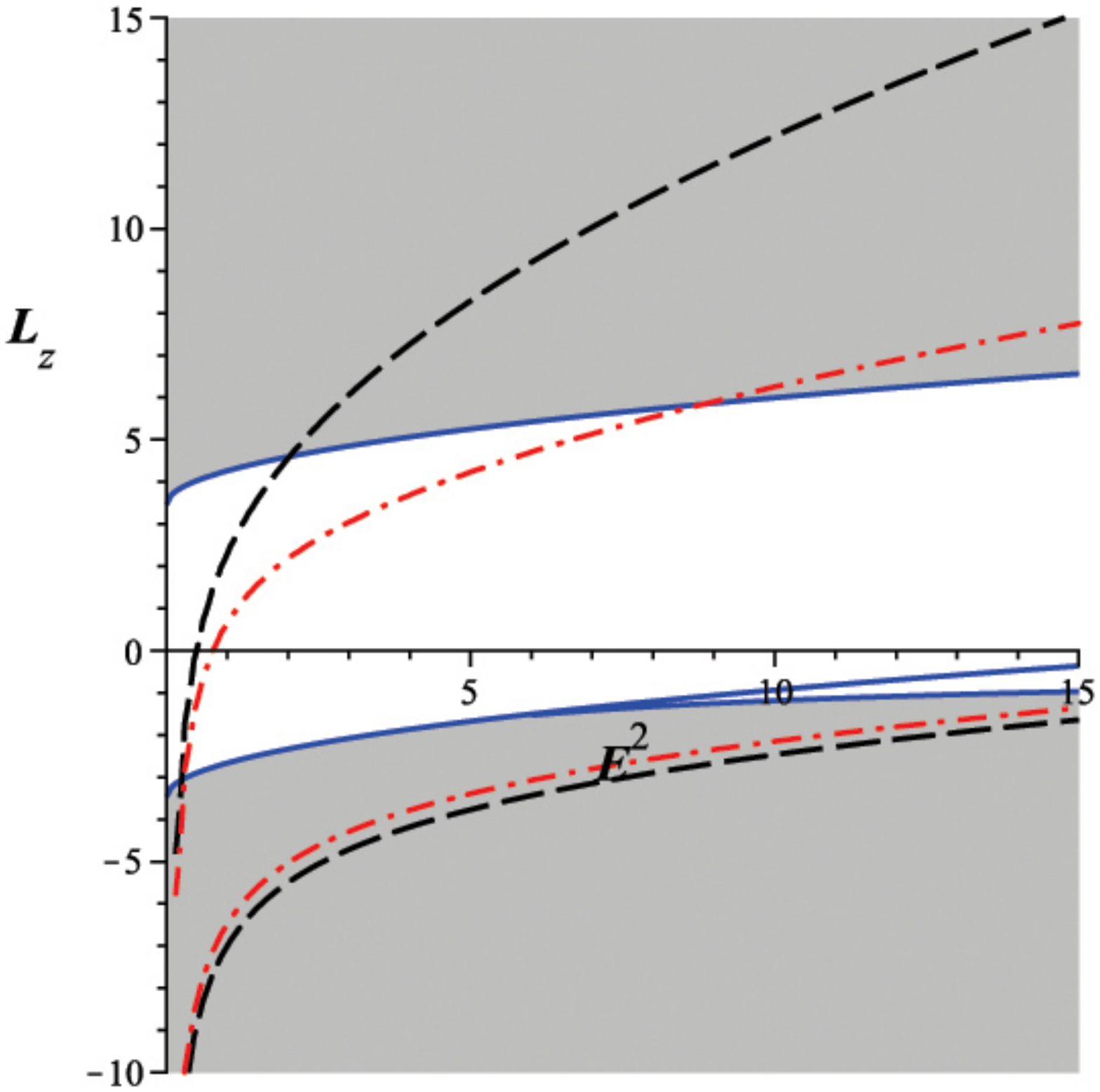}
}
\caption{ Combined parametric $L_z-E^2$ diagram for $r$ and $\theta$ motion of massless test particles with $a=0.8$, $M=1$, $Q=0.4$, $K=12$. Here solid lines represent the bounderies of $\theta$-motion, dotted-dashed lines represent bounderies of $r$-motion for KSBH while dashed lines represent bounderies of $r$-motion for KBH and shaded area depicts the forbidden region for geodesic motion of massless test particle.\label{fig:latitudinal_K_KS}}
\end{figure}
\noindent All the possible regions for geodesic motion of a massless test particle around KSBH are depicted in Fig.(\ref{fig:radial_motion_null}). 
Further the allowed regions for the motion of photons around KSBH and KBH are compared in Fig.(\ref{fig:latitudinal_K_KS}), which signifies clearly that the presence of charge in KSBH restricts the motion of a photon in radial direction in comparison to that of KBH. 
In other words, the allowed region for geodesic motion of massless test particle shrinks in the presence of the charge. 
An important point to notice here is that the motion of a photon is similar in two cases, as the presence of the charge has no effect on the motion of a photon in latitudinal direction.
\subsection{Types of Orbits}
List of all possible orbits for any arbitrary (massive or massless) test particle is given below \citep{Hackmann:2010tqa} (where $r_+$ and $r_-$ represent the event and cauchy horizon of the BH while $r_1$ and $r_2$ represent the real zeros of the radial function $\mathcal{R}(r)$):
\begin{itemize}
\item[(a)] {\it Transit Orbit} with r $\in$ ($-\infty,\infty$).
\item[(b)] {\it Escape Orbit} with r $\in$ [$r_1,\infty$) with $r_1>r_+$ or with r $\in$ ($-\infty,r_1,$] with $r_1<0$.
\item[(c)] {\it Two-World Escape Orbit} with [$r_1,\infty$] where $0<r_1<r_-$.
\item[(d)] {\it Crossover Two-World Escape Orbit} with [$r_1,\infty$) where $r_1<0$.
\item[(e)] {\it Bound Orbit} with r $\in$ [$r_1,r_2$] with $r_1,r_2>r_+$ or $0<r_1,r_2<r_-$.
\item[(f)] {\it Many-World Bound Orbit} with r $\in$ [$r_1,r_2$] where $0<r_1\leq r_-$ and $r_2\geq r_+$.
\item[(g)] {\it Terminating Orbit} with either r $\in$ [$0,\infty$) where $r_1\geq r_+$ or r $\in$ [$0,r_1$] with $0<r_1<r_-$.
\end{itemize}
\noindent As the presence of the {\it Terminating Orbits} require $r=0$ and $\theta=\pi/2$ simultaneously. 
In case of the geodesic motion of the photon around a KSBH, the condition for the presence of {\it Terminating Orbits} further restricts $Q=0$ which in turn implies the absence of the {\it Terminating Orbits} for photons around KSBH.
\begin{figure}[h!]
\centerline{
\includegraphics[scale=0.28]{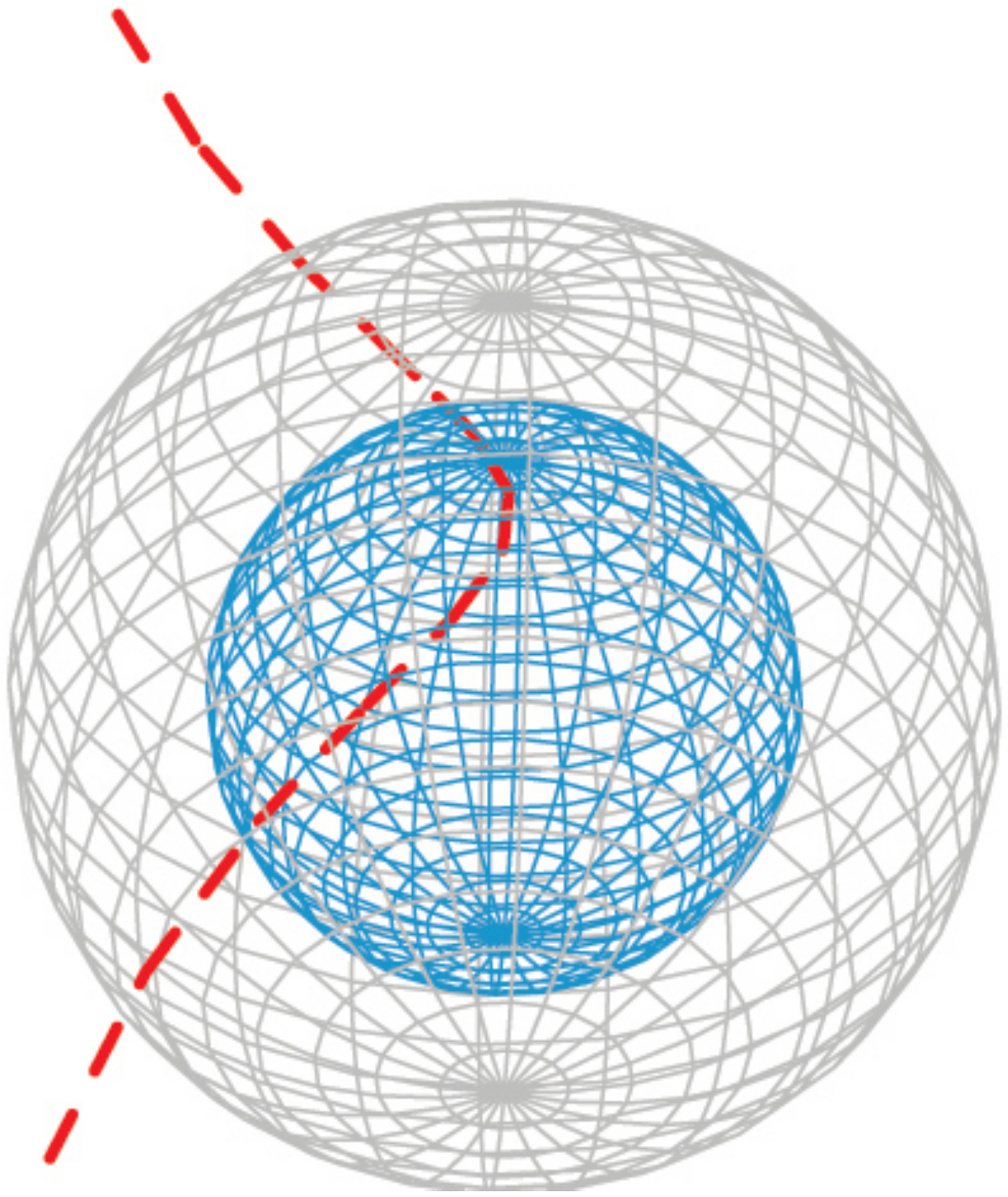}
\includegraphics[scale=0.23]{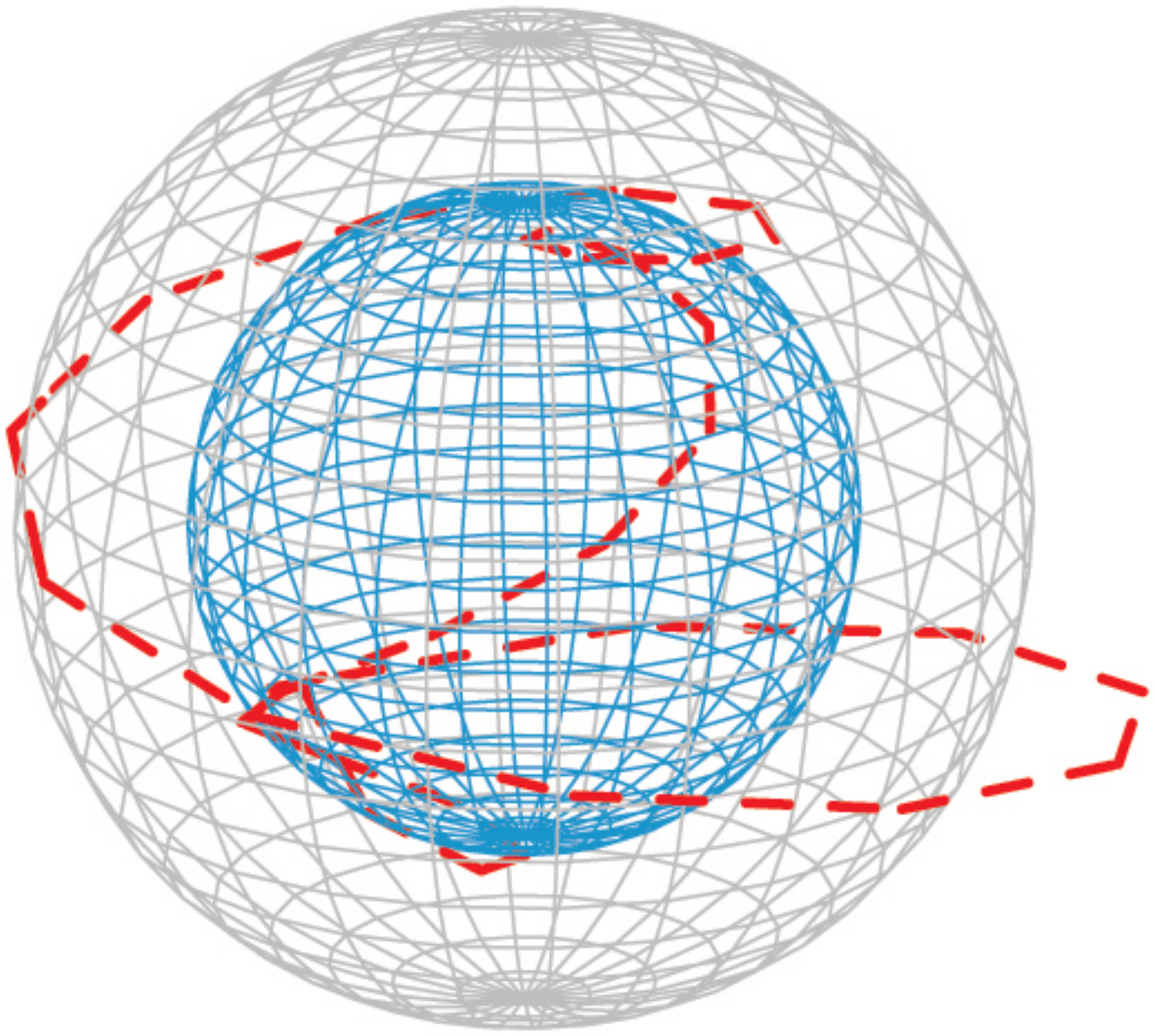}
\includegraphics[scale=0.27]{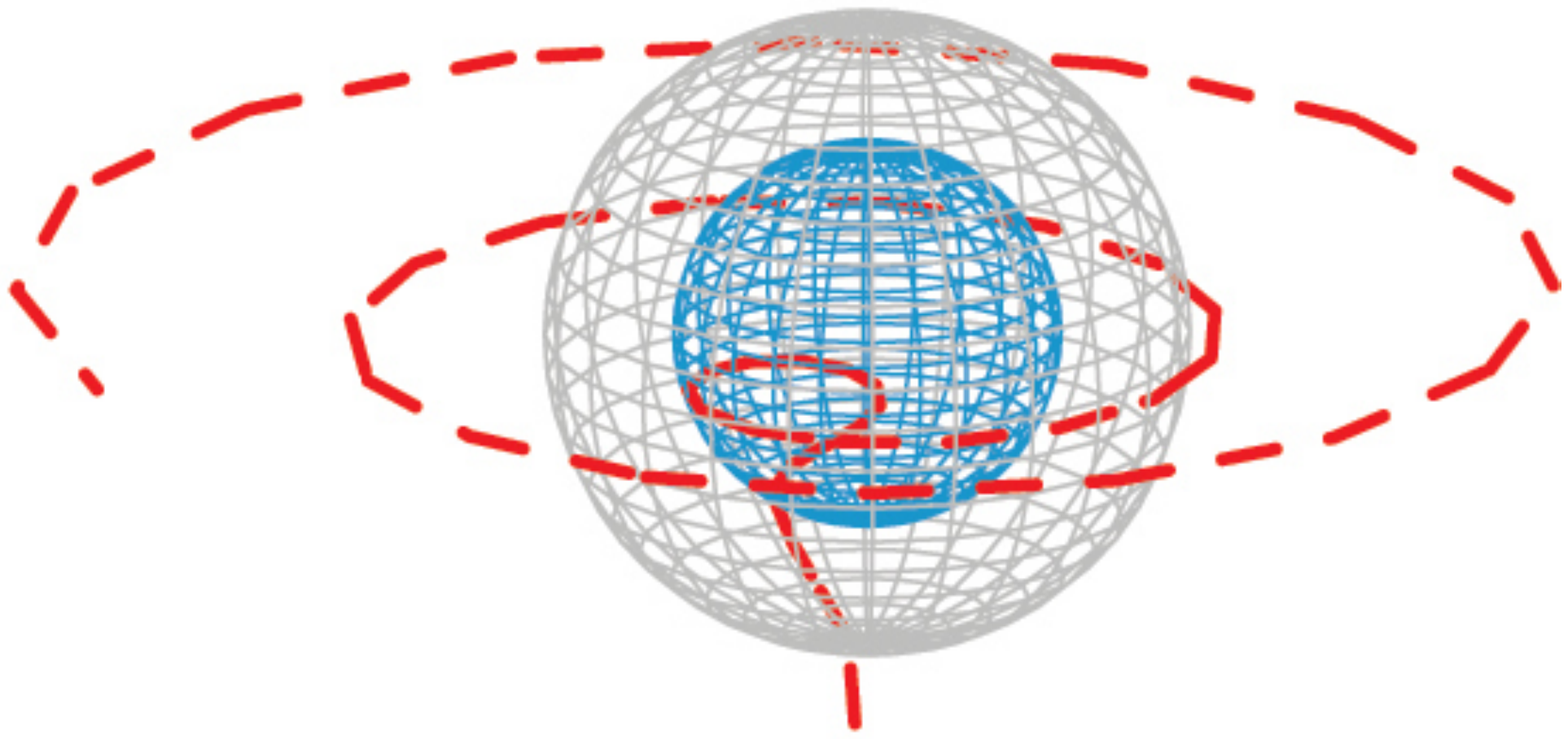}
}
\caption{(i)\emph{Left figure:} Transit orbit for a photon with $E = 4$, $L = -3$; (ii)\emph{Middle Figure:} Many-world bound orbit for a photon with $E = 2$, $L = 8$ and (iii)\emph{Right Figure:} Two-world escape orbit for photon with $E = 2$, $L = 2$ respectively with $M = 1$, $Q = 0.6$, $a = 0.8$, $K = 12$. Here the inner sphere represents the \emph{cauchy horizon} and the outer sphere represents the \emph{event horizon} of KSBH.}
\end{figure}
\section{Observable on angular plane}
\subsection{Bending of Light}

\noindent Using Eqs.(\ref{Rad}) and (\ref{phiequation}) alongwith the change of variable as $r+x=1/v$, the orbit equation for massless test particles (with $\delta = 0$) on angular plane (i.e. $\theta = \pi/2$) can be recasted as,
\begin{equation}
\left(\frac{dv}{d\phi}\right)^2=\left[j^2\left(\,(1-xv)^2+(1-xv)\,a^2\,v^2+2\,M\,a^2\,v^3\right)-4\,j\,M\,a\,v^3-\mathcal{B}(v)\,v^2\right]
\nonumber
\end{equation}
\begin{equation}
\left(\mathcal{B}(v)+2\,M\,x\,v^2+a^2\,v^2\right)\left(\mathcal{B}(v)+a^2\,v^2\right)/\mathcal{B}^2(v),
\label{eq:orbit-null}
\end{equation}
\vspace{2mm}
where,
$j=a/M$, $x=Q^2/M$ and $\mathcal{B}(u)=1-(x+2M)\,v$.
Expanding the right hand side of the Eq.(\ref{eq:orbit-null}) upto third order in $v$ as given below,
\begin{eqnarray}
\left(\frac{dv}{d\phi}\right)^2=j^2-2\,j^2\,x\,v+\left[j^2\left(x^2+2\,M\,x+3\,a^2\right)-1\right]v^2
\nonumber\\
+\left[j^2\left(a^2\,(2M-5)-4\,M\,x^2\right)-4\,j\,M\,a+y\left(1-2\,(a^2+Mx)\right)\right]\,v^3.
\label{eq:orbit-null-third-order}
\end{eqnarray}
Eq.(\ref{eq:orbit-null-third-order}) can further be re-expressed in the following form,
\begin{eqnarray}
\frac{d\phi}{dv}=\frac{1}{\sqrt{j^2+\mathcal{G}v+\mathcal{H}v^2+\mathcal{I}v^3}}\hspace{1mm}
\Rightarrow\hspace{1mm} \phi-\phi_0=\int^v_{v_0}\frac{dv}{\sqrt{j^2+\mathcal{G}v+\mathcal{H}v^2+\mathcal{I}v^3}}\hspace{0.5mm}.
\label{eq:int-bol}
\end{eqnarray}
Here $\mathcal{G}=-2j^2x$, $\mathcal{H}=j^2(x^2+2Mx+3a^2)-1$ and $\mathcal{I}=j^2\left[a^2(2M-5)-4Mx^2\right]-4a^2+y\left[1-2(a^2+Mx)\right]$. 
On integrating Eq.(\ref{eq:int-bol}), bending of light can be obtained as,
\begin{equation}
\Delta\phi=\frac{C}{\mathcal{N}}\arctan\left[\frac{\mathcal{N}v+j^2x}{\mathcal{N}\sqrt{j^2-2j^2xv-\mathcal{N}v^2}}\right]\,,
\label{eq:bol}
\end{equation}
with $\mathcal{N} = \left[1-j^2(2Q^2+3a^2)-j^2x^2\right]^{1/2}$. 
Here only the terms upto second order in $v$ are taken into the account in the right hand side of Eq.(\ref{eq:int-bol}).
\subsection{Perihelion precession}
\noindent Perihelion precession is calculated for timelike geodesics in
the equatorial plane (i.e. $\theta$ = $\pi$/2) since the angular momentum is conserved only in this plane, which is perpendicular to the rotation axis. For particle motion off the equatorial plane, the angular momentum would no longer be used as a conserved quantity.
Using 
\noindent Using Eqs.(\ref{Rad}) and (\ref{phiequation}) alongwith the change of variable as $r+x=1/v$, the orbit equation now be transcribed as (with $\delta = -1$),
\begin{equation}
\left(\frac{dv}{d\phi}\right)^2\,\,=\,\,\left(\mathcal{B}(v)+2\,M\,x\,v^2+a^2\,v^2\right)\left(\mathcal{B}(v)+a^2\,v^2\right)\times\frac{1}{\mathcal{B}^2(v)}\times
\nonumber
\end{equation}
\begin{equation}
\left[j^2\left((1-xv)^2+(1-xv)\,a^2\,v^2+2\,M\,a^2\,v^3\right)-4\,j\,M\,a\,v^3-\mathcal{B}(v)\,v^2-(\mathcal{B}(v)+a^2\,v^2)\frac{(1-xu)}{L_z^2}\right]
\label{eq:orbit-timelike}
\end{equation}
\vspace{2mm}
\noindent where all the parameters have their usual meanings as mentioned earlier.
Expanding the right hand side of Eq.(\ref{eq:orbit-timelike}) upto third order in $v$ as,
\begin{equation}
\left(\frac{dv}{d\phi}\right)^2=
\left[j^2\,a^2(2M-x)+y-4\,j\,M\,a+\frac{xa^2}{L_z^2}+2\,(Mx+a^2)\left(-2j^2\,x+\frac{x+y}{L_z^2}\right)-2\,y\,(a^2+Mx)\right]v^3
\nonumber
\end{equation}
\begin{equation}
+\left[-1+j^2\,(x^2+a^2)-\frac{x\,y+a^2}{L_z^2}
+2\,(M\,x+a^2)\left(j^2-\frac{1}{L_z^2}\right)\right]v^2
\nonumber
\end{equation}
\begin{equation}
+\left(-2\,j^2\,x+\frac{x+y}{L_z^2}\right)\,v+j^2-\frac{1}{L_z^2},
\label{eq:orbit-timelike-v3}
\end{equation}
with $y=x+2M$.
Differentiating Eq.(\ref{eq:orbit-timelike-v3}) w.r.t. $\phi$,
\begin{equation}
v^{\prime\prime}+\left[1-j^2(x^2+a^2)+\frac{xy+a^2}{L_z^2}-2(Mx+a^2)\left(j^2-\frac{1}{L_z^2}\right)\right]v=-j^2x+\frac{x+y}{2L_z^2}+
\nonumber
\end{equation}
\begin{equation}
\frac{3}{2}\left[j^2\,a^2\,(2M-x)+y-4\,j\,M\,a+\frac{xa^2}{L_z^2}+2(Mx+a^2)\left(-2j^2x+\frac{x+y}{L_z^2}\right)-2y(a^2+Mx)\right]v^2.
\label{eq:orbit-timelike-second-order}
\end{equation}
\vspace{2mm}\\
Now the Eq.(\ref{eq:orbit-timelike-second-order}) is of the form as presented in \citep{Damour:1994zq}.
On comparison, it leads the values of parameters as below,
\begin{equation}
\epsilon_1=-j^2(x^2+a^2)+\frac{xy+a^2}{L_z^2}-2(Mx+a^2)\left(j^2-\frac{1}{L_z^2}\right),
\end{equation}
\begin{equation}
A=-j^2x+\frac{x+y}{2L_z^2},
\end{equation}
\begin{equation}
\epsilon=\frac{3}{2}\left[j^2a^2(2M-x)+y-4jMa+\frac{xa^2}{L_z^2}+2(Mx+a^2)\left(-2j^2x+\frac{x+y}{L_z^2}\right)-2y(a^2+Mx)\right].
\end{equation}
\vspace{2mm}\\
With the assumption $a/L<<1$, the perihelion precession can be given as,
\begin{equation}
\Delta\phi=3\pi\left[j^2a^2(2M-x)+y-4jMa+\frac{xa^2}{L_z^2}+2(Mx+a^2)\left(-2j^2x+\frac{x+y}{L_z^2}\right)-2y(a^2+Mx)\right]
\nonumber
\end{equation}
\begin{equation}
-\pi\left[-j^2(x^2+a^2)+\frac{xy+a^2}{L_z^2}-2(Mx+a^2)\left(j^2-\frac{1}{L_z^2}\right)\right].
\label{eq:perihelion-precession}
\end{equation}
\begin{figure}[h!]
\centerline{
\includegraphics[scale=0.4]{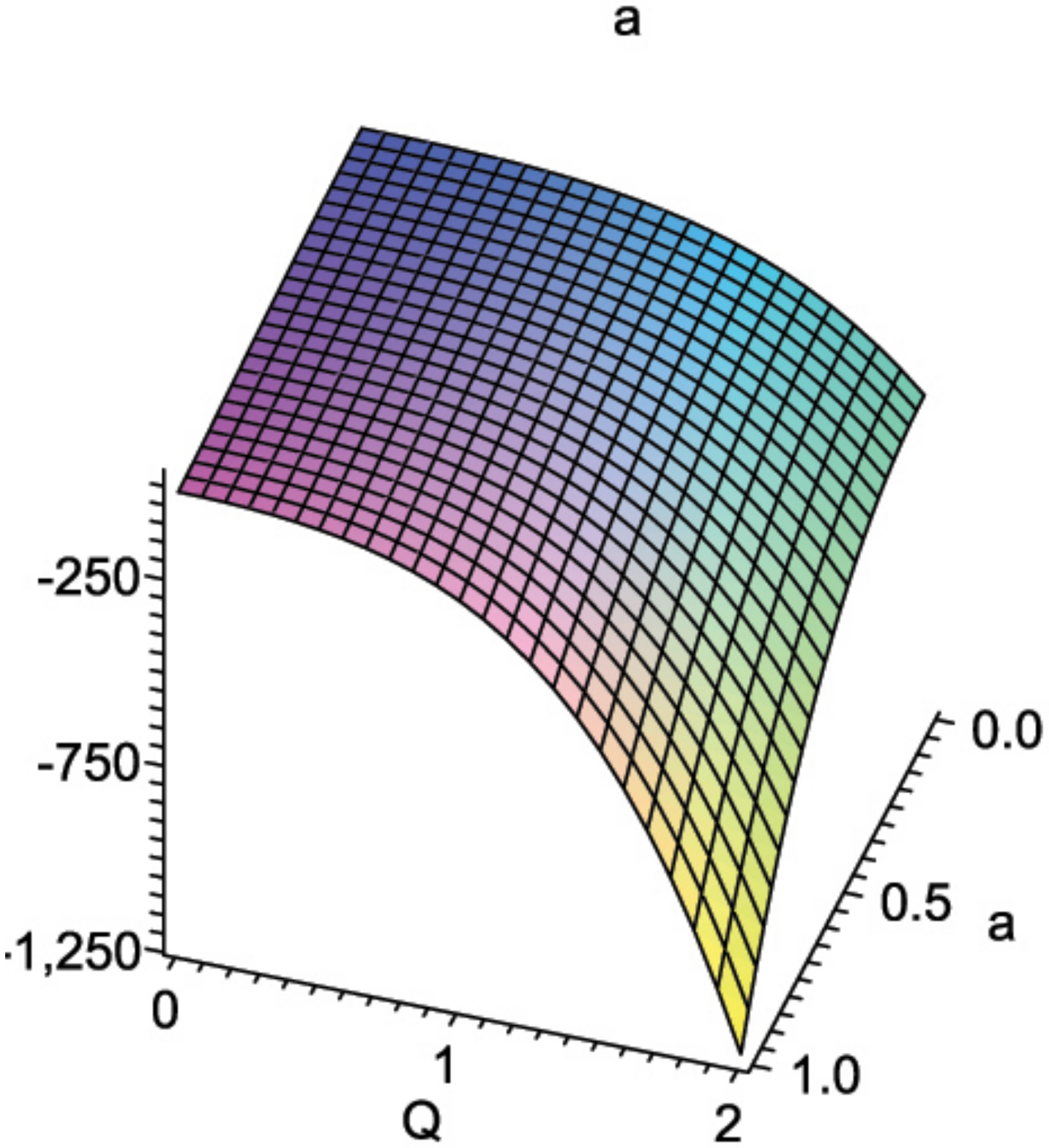}
\includegraphics[scale=0.4]{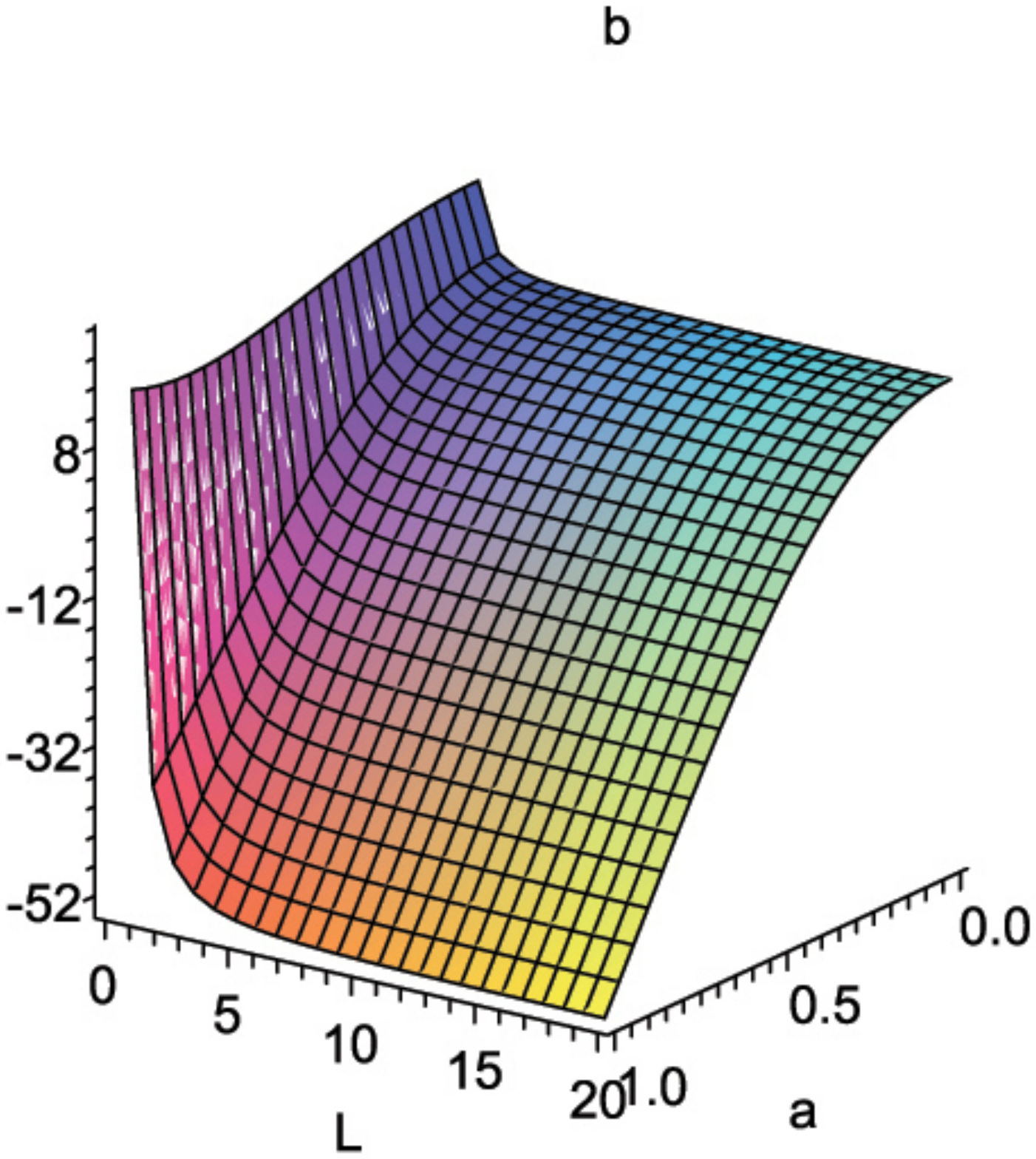}
}
\caption{\label{fig:P-shift-3d-Q-a}Effect of variation in: (a) charge parameter $Q$ and rotation parameter $a$ and (b) angular momentum of the incoming test particle $L_z$, on the perihelion shift of timelike geodesics, here vertical axis corresponds to the perihelion shift $\Delta\phi$.}
\end{figure}

\noindent Fig.\ref{fig:P-shift-3d-Q-a}(a) depicts the variation in the angle of precession ($\Delta\phi$) with BH charge $Q$ and specific angular momentum $a$ while $E$ and $L$ have fixed values. 
It is observed that an increment in the values of $Q$ or $a$ or both, results in the decline of the corresponding value of $\Delta\phi$. 
Hence precession of the particle weakens on increasing $Q$ and $a$. However, Fig.\ref{fig:P-shift-3d-Q-a}(b) depicts the variation in $\Delta\phi$ for fixed $Q$ and $E$. 
A small increment in the value of $L$ shows a sharp decline in the corresponding value of $\Delta\phi$ for fixed $a$ while the effect of the variation of $a$ is similar to that of Fig.\ref{fig:P-shift-3d-Q-a}(a).
\subsection{Red/blue shift of photons emitted by massive test particles following geodesics}
In the present section we have calculated red/blue shifts experienced by photons travelling towards a distant observer and emitted by massive particles following geodesics. 
For this purpose we have followed the approach as opted in \citep{Herrera-Aguilar:2015kea} and \citep{Becerril:2016qxf}.\\
\noindent The generalised expression for the frequency of a photon measured by an observer with proper $4$-velocity $U^{\mu}_C$ at point $P_C$ is
\begin{equation}
\omega_C=-k_\mu\,U^{\mu}_C|_{P_C},
\end{equation}
where the index $C$ refers to the emission (e) and/or detection
(d) at the corresponding space-time point $P_C$. 
The 4-velocities of the emitter and the detector
respectively are:
\begin{equation}
U^{\mu}_e\,=\,\left(U^t, U^r, U^\theta, U^\phi\right)|_e,\,\,\,\,\,\,and\,\,\,\,\,\,\,\,U^{\mu}_d\,=\,\left(U^t, U^r, U^\theta, U^\phi\right)|_d.
\end{equation}
If the detector is located far enough from the source, one can consider that the position of the observer at infinity ($r\rightarrow\infty$), which further gives the $4$-velocity as:
\begin{equation}
U^{\mu}_d\,=\,(1, 0, 0, 0),
\end{equation}
as $U^r_d$, $U^{\theta}_d$ and $U^{\phi}_d$ vanish, while $U^t_d\rightarrow E$ (i.e. $E=1$). 
On the other hand, a photon which is emitted or detected at point $P_C$ possesses a four momentum $k^{\mu}_C$ = $(k^t, k^r, k^{\theta}, k^{\phi})|_c$. 
The frequency shift associated with the emission and detection of photons is therefore,
\begin{equation}
1+z\,=\,\frac{\omega_e}{\omega_d}\,=\,\frac{\left(E_{\gamma}U^t-L_{\gamma}U^{\phi}-g_{rr}U^{r}k^r-g_{\theta\theta}U^{\theta}k^{\theta}\right)|_{e}}{\left(E_{\gamma}U^t-L_{\gamma}U^{\phi}-g_{rr}U^{r}k^r-g_{\theta\theta}U^{\theta}k^{\theta}\right)|_{d}}.
\end{equation}
The conditions for the circular equatorial (i.e. $V_{eff}(r)=0, V_{eff}^\prime(r)=0 $) bound orbits (i.e. $E<1$ and $K\geq0$) for massive test particles \citep{Bardeen:1972fi} gives the following condition for $r$ in case of KSBH,
\begin{equation}
r\,>\,r_{mb}\,=\,2M-\frac{x}{2}-a\pm\frac{1}{2}\sqrt{(4M-x)(4M-x-4a)},
\label{eq:marginally-bound-radius-timelike}
\end{equation}
where $r_{mb}$ represents the radius of marginally bound equatorial circular orbit for massive test particle around KSBH. Further condition for these above orbits following constraint given in Eq.(\ref{eq:marginally-bound-radius-timelike}) has another conditions to be a \emph{stable} orbit also i.e. $V_{eff}^{\prime\prime}(r)<0$,
\begin{equation}
r\,>\,r_{ms}\,=\,\left(M-\frac{x}{4}\right)\left(3+\mathcal{V}_2\,\mp\,\sqrt{(3-\mathcal{V}_1)(3+\mathcal{V}_1+2\mathcal{V}_2)}\right),
\label{eq:marginally-stable-radius-timelike}
\end{equation}
where
\begin{equation}
\mathcal{V}_1\,=\,1+\left(1-\frac{a^2}{\left(M-\frac{x}{4}\right)^2}\right)^{1/3}\left(\left(1+\frac{a}{\left(M-\frac{x}{4}\right)}\right)^{1/3}+\left(1-\frac{a}{\left(M-\frac{x}{4}\right)}\right)^{1/3}\right),
\nonumber
\end{equation}
\begin{equation}
\mathcal{V}_2\,=\,\sqrt{\frac{48{a^2}}{(4M-x)^2}\,+\,\mathcal{V}_1^2}.
\label{eq:ms-radius-functions}
\end{equation}
Here $r_{ms}$ represents the radius of marginally stable circular equatorial orbits for massive test particles around KSBH. Further, for circular and equatorial orbits, the expression for red/blue shift reduces to,
\begin{equation}
1+z\,=\,\frac{\left(E_{\gamma}U^t-L_{\gamma}U^{\phi}\right)|_{e}}{\left(E_{\gamma}U^t-L_{\gamma}U^{\phi}\right)|_{d}}\,=\,\frac{U^t_e-b_e{U^{\phi}_e}}{U^t_d-{b_d}{U^{\phi}_d}},
\end{equation}
where $b\equiv\frac{L_{\gamma}}{E_{\gamma}}$ is the apparent impact parameter and $U^r = U^{\theta} = 0$ for circular equatorial orbits.
\begin{equation}
b_{\pm}=-\frac{g_{t\phi}\pm\sqrt{g_{t\phi}^2-g_{tt}g_{\phi\phi}}}{g_{tt}},
\label{eq:impact-parameter-general}
\end{equation}
where $b_{-}$ and $b_{+}$ give rise to two
different shifts $z_1$ and $z_2$ of the emitted photons corresponding to a receding and to an approaching object
with respect to a very distant observer. 
$b_{-}$ and $b_{+}$ can either be evaluated at the emitter or detector position, since this quantity is preserved along the null geodesics i.e. $b_e$ = $b_d$. These different redshifts are given as:
\begin{equation}
{z_1}=\frac{U^t_{e}U^{\phi}_{d}b_{d_{-}}-U^t_{d}U^{\phi}_{e}b_{e_{-}}}{U^t_d\left(U^t_{d}-U^{\phi}_{d}b_{d_{-}}\right)},
\label{eq:blue_shift_01}
\end{equation}
\begin{equation}
{z_2}=\frac{U^t_{e}U^{\phi}_{d}b_{d_{+}}-U^t_{d}U^{\phi}_{e}b_{e_{+}}}{U^t_d\left(U^t_{d}-U^{\phi}_{d}b_{d_{+}}\right)}.
\label{eq:red_shift_01}
\end{equation}
Generally $|z_1|\neq|z_2|$, as the bending of light experienced
by the emitted photons on either side of the geometrical
center of the source. 
The differential rotation experienced by the detector can be explained by the terms $U^{\phi}_d$ and $U^t_d$. 
While the second term in the denominator of Eqs.(\ref{eq:blue_shift_01}) and (\ref{eq:red_shift_01}) resembles the contribution of the movement of the detector's inertial frame. 
If  $U^{\phi}_{d}$<< $U^t_{d}$, then the detector can be considered static at spatial infinity. 
The angular velocity of a distant detector from source can now be defined as,
\begin{equation}
\frac{U^{\phi}_{d}}{U^t_{d}}\,=\,\frac{d\phi}{dt}\equiv\Omega_d,
\end{equation}
Thus, when this quantity is
small i.e. $\Omega_d$ $<<$ $1$, the detector can be treated as static,
neglecting its relative movement. In terms of $\Omega_d$, the $z_1$ and $z_2$ reads as,
\begin{equation}
{z_1}=\frac{U^t_{e}\Omega_{d}b_{d_{-}}-U^{\phi}_{e}b_{e_{-}}}{U^t_d\left(1-\Omega_{d}b_{d_{-}}\right)},
\label{eq:blue_shift_02}
\end{equation}
\begin{equation}
{z_2}=\frac{U^t_{e}\Omega_{d}b_{d_{+}}-U^{\phi}_{e}b_{e_{+}}}{U^t_d\left(1-\Omega_{d}b_{d_{+}}\right)}.
\label{eq:red_shift_02}
\end{equation}
\noindent In order to have a close expression for the gravitational red/blue shifts experienced by the emitted photons one has to express the required quantities in terms of the KSBH parameters. Thus, the $U^{\phi}$ and $U^t$ components of
the 4-velocity for circular equatorial orbits are:
\begin{equation}
U^{\phi}(r,\pi/2)\,=\,\frac{\left(2Ma\right)E\,+\,\left(r+x-2M\right)L}{\left(r+x\right)\left(r^2+rx+a^2-2Mr\right)},
\label{eq:u-phi-equatorial}
\end{equation}
\begin{equation}
U^t(r,\pi/2)\,=\,\frac{\left(r\left(r+x\right)^2+a^2\left(r+x\right)+2Ma^2\right)E-\left(2Ma\right)L}{\left(r+x\right)\left(r^2+rx+a^2-2Mr\right)},
\label{eq:u-t-equatorial}
\end{equation}
here constants of motion $E$ and $L$ are given by,
\begin{equation}
E=\frac{r^{1/2}\left((r+x)(r-4M)+4M^2\right)^{1/2}\pm\,aM^{1/2}}{r^{3/4}(r+x)^{3/4}\,P^{1/2}},
\label{eq:E-circular-equatorial}
\end{equation}
\begin{equation}
L=\frac{\pm\,M^{1/2}\left(r(r+x)\mp2aM^{1/2}r^{1/2}+a^2\right)}{r^{3/4}(r+x)^{3/4}\,P^{1/2}},
\label{eq:L-circular-equatorial}
\end{equation}
with,
\begin{equation}
P\,=\,r^{1/2}\left((r+x)(r-6M)+9M^2\right)^{1/2}\pm\,2aM^{1/2},
\label{eq:polynomial}
\end{equation}
where the $\pm$ signs again correspond to the co-rotating
and counter-rotating objects (either the emitter or the
detector) with respect to the direction of the angular velocity
of the BH \citep{Bardeen:1972fi}. 
On substituting the corresponding values of $E$ and $L$ from Eqs.(\ref{eq:E-circular-equatorial}) and (\ref{eq:L-circular-equatorial}) into Eqs.(\ref{eq:u-phi-equatorial}) and (\ref{eq:u-t-equatorial}), one obtains
\begin{equation}
U^{\phi}(r,\pi/2)\,=\,\frac{\pm\,M^{1/2}}{r^{3/4}(r+x)^{3/4}\,P^{1/2}},
\label{eq:u-phi-E-L}
\end{equation}
\begin{equation}
U^t(r,\pi/2)\,=\,\frac{r^{3/4}(r+x)^{3/4}\pm\,aM^{1/2}}{r^{3/4}(r+x)^{3/4}\,P^{1/2}}.
\label{eq:u-t-E-L}
\end{equation}
where $P$ is defined in the Eq.(\ref{eq:polynomial}). Hence the angular velocity of a source orbiting in a circular equatorial orbit around the KSBH will be,
\begin{equation}
\Omega_{\pm}\,=\,\frac{\pm\,M^{1/2}}{r^{3/4}(r+x)^{3/4}\pm\,aM^{1/2}}
\label{eq:angular-velocity}
\end{equation}
this angular velocity corresponds to either the
emitter or the detector of photons, in which case the subscripts
$e$ and $d$ must be used respectively.\\
\noindent On the other side, for the KSBH metric, Eq.(\ref{eq:impact-parameter-general}) gives the following expression for the impact parameter $b(r)$ which explains the gravitational bending of light for photons travelling in circular and equatorial orbits,
\begin{equation}
b_{\pm}\,=\,\frac{-2aM\pm\,(r+x)\sqrt{r(r+x)+a^2-2Mr}}{r+x-2M}.
\label{eq:impact-parameter-circular-equatorial}
\end{equation}
where the maximum character of $b(r)$ is considered, i.e. it is assumed as the photons are emitted at the point where $k^r$= 0.  Hence, for KSBH the expressions for red and blue shifts reduces to:
\begin{equation}
z_{red}\,=\,\pm\,M^{1/2}\frac{r^{3/8}_d(r_d+x)^{3/8}\,P^{1/2}_d}{r^{3/8}_e(r_e+x)^{3/8}\,P^{1/2}_e}\times\frac{\left(r^{3/4}_d(r_d+x)^{3/4}-r^{3/4}_e(r_e+x)^{3/4}\right)}{\left(r^{3/4}_d(r_d+x)^{3/4}\pm\,aM^{1/2}\right)}\times\frac{\mathcal{X_+}}{\mathcal{Y_+}},
\label{eq:red-shift-KSBH}
\end{equation}
\begin{equation}
z_{blue}\,=\,\pm\,M^{1/2}\frac{r^{3/8}_d(r_d+x)^{3/8}\,P^{1/2}_d}{r^{3/8}_e(r_e+x)^{3/8}\,P^{1/2}_e}\times\frac{\left(r^{3/4}_d(r_d+x)^{3/4}-r^{3/4}_e(r_e+x)^{3/4}\right)}{\left(r^{3/4}_d(r_d+x)^{3/4}\pm\,aM^{1/2}\right)}\times\frac{\mathcal{X_-}}{\mathcal{Y_-}},
\label{eq:blue-shift-KSBH}
\end{equation}
where
\begin{equation}
\mathcal{X_{\pm}}\,=\,2aM\pm(r_e+x)\sqrt{r_e(r_e+x)-2Mr_e+a^2},
\label{eq:X}
\end{equation}
\begin{equation}
\mathcal{Y_{\pm}}\,=\,r^{3/4}_d(r_d+x)^{3/4}(r_e+x-2M)\pm\,M^{1/2}(r_e+x)\left(a\pm\sqrt{r_e(r_e+x)-2Mr_e+a^2}\right).
\label{eq:Y}
\end{equation}
\vspace{2mm}
\noindent If detector is very distant from the source then the following condition is fulfilled $r_d$ >> $M$ $\geq$ $a$, the red and blue shifts respectively become,
\begin{equation}
z_{red}\,=\,\frac{\pm\,M^{1/2}\mathcal{X_+}}{r^{3/8}_e(r_e+x)^{3/8}(r_e+x-2M)\,P^{1/2}_e},
\label{eq:red-shift-far}
\end{equation}
\begin{equation}
z_{blue}\,=\,\frac{\pm\,M^{1/2}\mathcal{X_-}}{r^{3/8}_e(r_e+x)^{3/8}(r_e+x-2M)\,P^{1/2}_e}.
\label{eq:blue-shift-far}
\end{equation}
\noindent In this special case, the mass and rotation parameters of the KSBH can be obtained from the measured red and blue shifts of the photons emitted by the stars
through Eqs.(\ref{eq:red-shift-far}) and (\ref{eq:blue-shift-far}). Hence the closed expression for the
rotation and mass parameters for KSBH are given as,
\begin{equation}
a^2\,=\,\frac{1}{\mathcal{A}}\left(r_e\,(r_e+x)^2\,(r_e+x-2M)\,\alpha\right),
\label{eq:rotation-BH}
\end{equation}
and
\begin{equation}
\left(16(r_e+x)M^3-\mathcal{A}\,(r_e+x-2M)\,(r_e+x-3M)\right)^2\,=\,4\alpha\,M\,r_{e}(r_e+x)(r_e+x-2M)^3\,\mathcal{A}.
\label{eq:mass-KSBH}
\end{equation}
where $\mathcal{A}\,=\,4M^2\beta-(r_e+x)^2\alpha$, $\alpha=(z_{red}+z_{blue})^2$ and $\beta=(z_{red}-z_{blue})^2$. 
In order to obtain the exact value of mass parameter, Eq.(\ref{eq:mass-KSBH}) is needed to be solved numerically. 
In general, Eq.(\ref{eq:mass-KSBH}) can have maximum eight real values theoretically. To have a real solution corresponding to $M$ first constraint is Eq.(\ref{eq:polynomial})$>0$. Further as we have constrained the motion of massive test particles along bound circular equatorial orbits, Eq.(\ref{eq:marginally-bound-radius-timelike}) and Eq.(\ref{eq:marginally-stable-radius-timelike}) must be followed as well.
\section{Summary and Conclusions}
The geodesic motion and other closely related aspects in the background of KSBH arising in the heterotic string theory are investigated in detail in this paper. 
The metric of KSBH spacetime being axially symmetric, offers the diverse possibilities for the geodesic motion of arbitrary test particles around it.  
Physically allowed regions for the photon's motion in latitudinal as well as radial direction are discussed in detail alongwith the corresponding effective potential and orbit structure. 
As KSBH spacetime arises as the charged-dilaton generalisation of KBH spacetime, all the results obtained are found to be modified accordingly, except the latitudinal motion of the test particle around this BH. 
Since the governing function for the motion of the test particles in $\theta$-direction is independent of charge, the latitudinal motion of the test particles have similar properties to that of corresponding to KBH spacetime. 
However, the presence of charge and dilaton field is manifested in the difference arising in the radial motion of test particles. Different observables such as bending of light and precessional angle of perihelion are found to depend on these different parameters involved in. Some of the key results obtained are summarised below:
\begin{itemize}
\item[(i)] In view of the geodesic motion of the test particles, the nature of effective potential is discussed in radial as well as latitudinal direction.
\item[(ii)] A special class of spherical photon orbits is obtained along with the expression for the turning point for radial photons. 
Dependence of photon motion within this class of solution is discussed explicitly in view of the different BH parameters.
\item[(iii)] A detailed discussion is made on the allowed regions for geodesic motion of test particles around KSBH in more generalised way by including non-equatorial motion of the photons into the account.
\item[(iv)] The conditions for different types of possible orbits are discussed with specific values of parameters along with the corresponding orbit structure; such as, transit orbits, many worls bound orbits, two-world bound orbits etc. 
It is observed that no terminating orbits are possible for photons around KSBH due to presence of the BH charge which mimics as a dark energy parameter \citep{Nandan:2009kt, Nandan:2016ksb}.
\item[(v)] Observables on the angular plane (viz. bending of light and perihelion precession for massive test particles) are analysed as special cases.
\item[(vi)] The rotation and mass parameters for KSBH are calculated in terms of the red/blue shifts of the photons in circular and equatorial orbits emitted by the massive test particles.
\end{itemize}
In order to have mass and spin parameters of astrophysical BHs we hope to extensively work out the red/blue shift approach opted in present paper for rotating BHs with dark energy backgrounds. In addition to this, as it is reported that KSBH follows cosmic censorship conjecture \citep{Gwak:2016gwj}, it will be timely to look for gyromagnetic frequency and other effects related to KSBH. We intend to report on these issues in near future.
\section{Acknowledgement}
The authors acknowledge the support from IUCAA, Pune, under its
visitors program where a part of this work was carried out. 
One of the author (HN) would like to thank Department of Science and Technology, New Delhi for the financial support through grant no. SR/FTP/PS-31/2009.

\end{document}